\documentclass[prd,aps,superscriptaddress,nofootinbib,amsmath,amssymb,showpacs,9pt]{revtex4}
\usepackage{graphicx}% Include figure files
\usepackage{braket}
\usepackage{nicefrac}
\usepackage{float}%
\usepackage{amsmath}
\usepackage{hyperref}
\usepackage{multirow}
\usepackage{slashed}
 \usepackage{bbold} % for identity matrix

\usepackage[usenames, dvipsnames]{color}

\def\slashchar#1{\setbox0=\hbox{$#1$}
   \dimen0=\wd0 \setbox1=\hbox{/} \dimen1=\wd1
   \ifdim\dimen0>\dimen1 \rlap{\hbox to \dimen0{\hfil/\hfil}} #1
   \else  \rlap{\hbox to \dimen1{\hfil$#1$\hfil}} / \fi}

\begin{document}

\title{Charged current neutrino scattering from nucleons}

\author{M. Sajjad \surname{Athar}}
\email{sajathar@gmail.com}
\affiliation{Department of Physics, Aligarh Muslim University, Aligarh-202 002, India}
\author{A. \surname{Fatima}}
\affiliation{Department of Physics, Aligarh Muslim University, Aligarh-202 002, India}
\author{S. K. \surname{Singh}}
\affiliation{Department of Physics, Aligarh Muslim University, Aligarh-202 002, India}
\author{F. \surname{Zaidi}}
\affiliation{Department of Physics, Aligarh Muslim University, Aligarh-202 002, India}

\begin{abstract} 
In this work, we study the charged current induced neutrino and antineutrino scattering from the free nucleon target. 
This study has been performed in the energy range of a few GeV, relevant for the  (anti)neutrino oscillation experiments with accelerator and atmospheric neutrinos. 
For a few GeV neutrino, the contribution to the cross section mainly comes from the quasielastic, the inelastic production of mesons like pion, kaon, eta, and hyperons as well as from the deep inelastic scattering by the weak currents in $\Delta S$=0 and $\Delta S$=1 sectors.  
The numerical results are presented for the $Q^2$ distribution of the differential cross section for all the aforementioned processes.
The effect of the cut on the center of mass energy $W$ has been explicitly discussed. 
\end{abstract}
\maketitle

\section{Introduction}
\label{intro}

In the Standard Model of Particle Physics, the neutrino is an elementary, massless, electrically neutral particle with spin $\frac{1}{2}$, which interacts only via the weak interaction. Neutrinos come in three generations, each associated with a corresponding charged lepton, i.e., the electron neutrino ($\nu_e$), the muon neutrino ($\nu_\mu$), and the tau neutrino ($\nu_\tau$). Each generation of neutrino has an independent lepton number, which is conserved in all processes. Additionally, for each type of neutrino, there exists a corresponding antineutrino with an opposite lepton number. Neutrino physics has been a leading area of research for over seven decades, yet this elusive particle continues to pose significant challenges. The history of neutrino physics has been full of surprises, challenging our expectations about symmetry principles and conservation laws in nature. 

The establishment of neutrino oscillation phenomenon through many experiments performed in the last three decades using solar, accelerator, reactor and atmospheric neutrino/antineutrino beams tell us that the interacting neutrino flavor states $\nu_{e}$, $\nu_{\mu}$, and $\nu_{\tau}$, and the propagating mass states, generally represented by $\nu_{1}$, $\nu_{2}$, and $\nu_{3}$  are different.  The neutrino flavor states do not have well defined mass as they are not mass eigenstate but are mixture of three neutrino mass states $\nu_{1}$, $\nu_{2}$, and $\nu_{3}$, with masses $m_{1}$, $m_{2}$, and $m_{3}$, of which at least two are non-zero. 
 The physical states $\nu_{e}$, $\nu_{\mu}$, and $\nu_{\tau}$ are related with the mass states $\nu_{1}$, $\nu_{2}$, and $\nu_{3}$ through a mixing matrix known as PMNS~(Pontecorvo-Maki-Nakagawa-Sakata) matrix, which is described in terms of three mixing angles viz. $\theta_{12}$, $\theta_{13}$ and $\theta_{23}$ and a CP-violating phase $\delta$, and the oscillation probability for a neutrino of one flavor say $\nu_f$ to an another flavor say $\nu_{f^\prime}$, $f \ne f^\prime=e, \mu, \tau$ also depends on the difference of the mass square of the three mass eigenstates viz. $\Delta m_{12}^2$ and $\Delta m_{13}^2$.
 
  Currently, the different neutrino experiments, which are either ongoing or in development, aim at resolving some of the important properties of neutrinos or the astrophysical phenomenon where neutrinos can be an important source of information. These experiments have the potential to significantly advance our understanding in the following key areas:
\begin{itemize}
 \item {\bf Neutrino Mass Hierarchy:} Determining the precise order of neutrino masses—whether the mass states follow a normal hierarchy ($m_1 ~<~ m_2~ <~ m_3$) or an inverted hierarchy ($m_3 ~<~ m_1~ < ~m_2$) remains a critical unanswered question. Upcoming high-precision experiments like the Deep Underground Neutrino Experiment (DUNE)~\cite{DUNE:2018tke, DUNE:2024wvj} at the Fermilab in the United States, Hyper-Kamiokande~\cite{Hyper-Kamiokande:2018ofw} in Japan, The Jiangmen Underground Neutrino Observatory~(JUNO)~\cite{Cerrone:2024vvk} in China, etc., aim to resolve this by studying differences in the neutrino oscillation patterns.  

\item {\bf Absolute Neutrino Mass Scale:} Although we know neutrinos have mass from the  oscillation experiments, the absolute scale of these masses is still unknown. The KATRIN experiment~\cite{KATRIN:2022ayy} seeks to measure the electron neutrino's mass directly through the beta decay of tritium. However, the tiny mass of neutrinos makes this measurement extremely challenging, and determining the masses of other neutrino flavors is even more difficult as currently the measurements are being done in the pion and tauon decay experiments.

\item {\bf Nature of Neutrinos:} It is still unclear whether neutrinos are Dirac particles (distinct from their antiparticles and the lepton numbers for the particle and the antiparticle are different) or Majorana particles (identical to their antiparticles and no lepton number is assigned). 

\item  {\bf CP Violation in the Lepton Sector:} While CP violation (which breaks the symmetry between matter and antimatter) has been observed in the hadron sector, it has not yet been confirmed in the lepton sector. Recent studies have placed limits on the CP-violating phase $\delta$, but more precise experiments are needed. Discovering CP violation in neutrino oscillations could provide insights into why the universe is dominated by matter. Experiments like DUNE~\cite{DUNE:2018tke, DUNE:2024wvj} and Hyper-Kamiokande~\cite{Hyper-Kamiokande:2018ofw} are focused on measuring CP-violating phases in the neutrino oscillations.  

\item {\bf Sterile Neutrinos:} The LSND experiment~\cite{LSND:1996ubh, LSND:2001aii} at the Los Alamos using $\mu$-decay at rest and observing $\bar\nu_e$ appearance in $\bar\nu_\mu$, followed by the MiniBooNE~\cite{MiniBooNE:2007uho, MiniBooNE:2008yuf} accelerator experiment using $\nu_e$/$\bar\nu_e$ appearance in $\nu_\mu$/$\bar\nu_\mu$ beam at the Fermilab have observed excess of electron events which can not be explained in the standard three flavor neutrino oscillation scenario and hints towards a possible fourth flavor~(sterile) neutrino, which does not interact with the matter in the ordinary way like the other three neutrinos.  The recent experiments like the MicroBooNE~\cite{Denton:2021czb}, IceCube~\cite{IceCube:2024dlz}, STEREO~\cite{STEREO:2022nzk}, etc. have refuted the idea of sterile neutrino. Further investigations are necessary to confirm or deny the presence of sterile neutrinos.
 
\item {\bf T-Violation Studies:} In the Standard Model, CPT is an exact conservation law. Since CP is violated in weak interactions (such as neutral kaon regeneration and B-meson oscillation), T-violation should also occur. Experimental efforts are being made to study T-violation in the (anti)neutrino sector.

\item {\bf High-Energy Neutrino Astronomy:} The detection of high-energy neutrinos from the astrophysical sources by the experiments like the IceCube~\cite{IceCube:2024dlz}, has opened a new window into the universe. 
There are other experiments like PINGU~\cite{IceCube:2016xxt} and KM3NeT~\cite{Ferrara:2024try} to observe the high energy neutrinos. 
However, understanding the sources and mechanisms producing these neutrinos remains a significant challenge, both theoretically and experimentally. 
\end{itemize}

In the (anti)neutrino energy region of a few GeV to which some of the above issues may be addressed, experiments are being conducted or planned using nuclear targets like $^{12}$C, $^{16}$O, $^{40}$Ar, $^{56}$Fe, and $^{208}$Pb. 
These experiments measure events that are a convolution of the energy-dependent (anti)neutrino flux, neutrino-nucleus cross sections, detector efficiency and acceptance. 
The neutrino flux is determined by observing the charged leptons produced in the  quasielastic reaction between the neutrinos and the nuclear target. 
For 1 GeV neutrinos, the nuclear medium effect causes an energy smearing of about 150 MeV. 
To mitigate the effects of the nuclear medium, recent neutrino oscillation experiments use the same nuclear target for both the near and the far detectors. 
However, due to energy modulation and the neutrino flavor oscillation, it is expected that the effect of nuclear medium is not exactly the same at the near and the far detectors. 
Thus, a thorough understanding of the energy dependence of the neutrino-nucleon and the neutrino-nucleus cross sections is essential, particularly for the current neutrino oscillation experiments that use nuclear targets. 
Achieving an accuracy of $2-3\%$ in the systematic uncertainties requires a  comprehensive knowledge of these cross sections. 
However, presently it has been estimated that due to the lack in the understanding of the energy dependent neutrino-nucleus cross sections, there is an uncertainty of about $25-30\%$ in these systematics. 
Thus, precisely measuring neutrino interaction cross sections is crucial for interpreting the oscillation experiments and understanding the neutrino interactions with matter. Efforts are ongoing to improve our knowledge of the (anti)neutrino-nucleon and the (anti)neutrino-nucleus cross sections through both the experimental and the theoretical works.  

Studying the neutrino interactions with matter is not only vital for advancing neutrino physics but also for gaining deeper insights into the hadronic interactions in the weak  sector, where the additional contribution from the axial vector current exists alongside the vector current. 
Moreover, in the weak sector, the various vector and axial vector form factors appearing in the hadronic current are determined assuming  various symmetries like the time reversal invariance, SU(3) symmetry, conserved vector current~(CVC) hypothesis, partial conservation of the axial vector current~(PCAC) hypothesis etc., therefore, a better understanding of the neutrino-nucleon/nucleus cross sections is important in order to test these symmetry properties~\cite{SajjadAthar:2022pjt, Athar:2020kqn, 
SajjadAthar:2021prg, Ruso:2022qes}. 
% {\bf Note: Precisely determining neutrino oscillation parameters.  Contributing to the quest for understanding CP violation and the neutrino mass hierarchy: Already discussed in earlier sections}
     
Historically, the theory of weak interaction was first formulated by Fermi~\cite{Fermi:1934sk}, and independently by Perrin~\cite{Perrin:1933} for the nuclear $\beta$ decay  with no change of total spin $J$ of the parent nucleus, i.e., $\Delta J = 0$. 
The interaction Lagrangian was taken to be of the current-current interaction type, in which the vector charged currents $l_{\mu}$ and $j^{\mu}$ constructed from the lepton pairs $(e^{-}, \bar{\nu}_{e})$ and nucleon pairs $(n,p)$, i.e. $l_{\mu} = \bar{\Psi}_{\nu} \gamma_{\mu} \Psi_{e}$ and $j^{\mu} = \bar{\Psi}_{n} \gamma^{\mu} \Psi_{p}$ were assumed to interact with strength $G$.  
The theory was extended by Gamow and Teller~\cite{Gamow:1936jk} to include the nuclear $\beta$ decays with $|\Delta J| = 1$, by adding an additional term of the axial vector current-current interaction. 
% The Lagrangian were able to describe the main features of the allowed Fermi and Gamow-Teller $\beta$ decays of nuclei. 

With the theoretical prediction~\cite{Lee:1956vjd, Lee:1956qn} and the experimental observation of parity
violation~\cite{Wu:1957my} in $\beta$ decays, the experimental determination of the neutrino helicity to be $-1$~\cite{Goldhaber:1958nb}, and the longitudinal polarization of $e^{-}~(e^{+})$ in $\beta^{-}~(\beta^{+})$ decays to be $\mp \beta$~\cite{Frauenfelder:1957na,  Page:1957zza}, the $V-A$ theory of weak interactions was proposed by Sudarshan and Marshak~\cite{Sudarshan:1958vf}, Feynman and Gell-Mann~\cite{Feynman:1958ty}, and Sakurai~\cite{Sakurai:1958zz}, using chiral~($\gamma_5$) invariance~\cite{Salam:1957st, Landau:1957tp, Lee:1957qr}, in which the interaction Lagrangian is given by~\cite{Athar:2020kqn}:
\begin{eqnarray}\label{eq:L_int}
 {\cal L}_{int}^{Weak} = \frac{G}{\sqrt{2}} l_{\mu} J^{\mu},
\end{eqnarray}
where 
\begin{eqnarray}
 l_{\mu} &=& \bar{\Psi}_{e} \gamma_{\mu}(1-\gamma_{5}) \Psi_{\nu}, \\
 J^{\mu}&=& \bar{\Psi}_{p} \gamma^{\mu} \left(C_{V} -C_{A} \gamma_{5} \right) \Psi_{n}.
\end{eqnarray}

The $V-A$ theory of weak interaction was able to explain all the available data on the nuclear and nucleon $\beta$ decays and low energy neutrino-nucleus scattering processes induced by the charged currents $J^{\mu}$ in the lowest order of perturbation theory, with $GC_{V} = G_{F}$ and $C_{A}/C_{V} = 1.26$. 
Later on, the theory was extended by Cabibbo~\cite{Cabibbo:1963yz} to describe the weak interactions of strange particles by adding a piece involving the interactions of the $\Delta S=1$ hadronic current $J_{\Delta S=1}^{\mu}$, which interacted with the leptonic current with a strength weaker than that of  the $\Delta S=0$ current of the $\beta$ decays and writing the hadronic current in Eq.~(\ref{eq:L_int}) as
\begin{equation}
 J^{\mu} = \cos \theta_{C} J^{\mu}_{\Delta S=0} + \sin \theta_{C} J^{\mu}_{\Delta S=1}
\end{equation}
and identifying $GC_{V} = G_F\cos\theta_{C}$, where $\theta_C$ is the Cabibbo angle, which is phenomenologically determined to be $13.1^o$ by studying the weak interaction of the strange particles.
 
The $V-A$ theory gave divergent results for the cross section even in the case of the simple process of $\nu_{e}e^{-}$ scattering and seemed to be non-renormalizable. 
The divergence of the local $V-A$ theory was tried to be cured by introducing an intermediate vector boson~(IVB) $W$, which has to be massive, unlike the photon in the case of electromagnetic interaction, in order to reproduce a local interaction in the low energy limit. 
% The charged IVB interacts with the lepton current and the hadron current, each with a strength $g_{W}$ and the $V-A$ interaction is recovered in the low energy limit of the interaction diagram given in Fig.~\ref{}, with $g_{W}^{2} = \frac{G_{F} M_{W}^{2}}{\sqrt{2}}$. 
The introduction of IVB helped only to postpone the divergence to the higher energies and did not prevent the occurrence of divergences in the higher order perturbation theory. The large mass of IVB mediating the weak interaction and zero mass of photon mediating the electromagnetic interaction was a great obstacle in unifying the weak and electromagnetic interactions. 
The attempts to formulate a renormalizable theory of weak interaction which also unifies the electromagnetic and weak interactions led to the formulation of the standard model~\cite{Weinberg:1967tq, Salam:1968rm} of the electroweak interactions. 

In the standard model, the interaction Lagrangian for the weak charged current~(CC) and neutral current~(NC) are written as~\cite{Athar:2020kqn}:
\begin{eqnarray}
 {\cal L}_{int}^{CC} &=& -\frac{g}{2\sqrt{2}}\sum_{l=e,\mu,\tau}\left[ \overline{\nu}_{l}\gamma^{\mu}(1-\gamma_5)l W_{\mu}^{+}+h.c.
\right],\\
 {\cal L}_{int}^{NC} &=& -\frac{g}{2\cos\theta_W}\sum_{l=e,\mu,\tau}\left[\overline{\nu}_{l}\gamma^{\mu} (g_{V}^{\nu_{l}} - 
g_{A}^{\nu_{l}} \gamma_{5})\nu_{l} 
  +\bar{l}\gamma^\mu (g_V^l -g_A^l \gamma_5)l \right] Z_\mu ,
\end{eqnarray}where
\begin{eqnarray}\label{gve_ref}
g_V^l = 2 \sin^2\theta_W -\frac{1}{2} , \qquad\qquad g_V^{\nu_{l}} = \frac{1}{2}, \qquad \qquad
g_A^l = -\frac{1}{2}, \qquad \qquad g_A^{\nu_{l}} = \frac{1}{2},
\end{eqnarray}
with $\theta_{W}$ being the Weinberg angle.
\begin{table}[htbp]
\begin{center}
 \begin{tabular}{|c|c|c|}\hline
States&$g_V$&$g_A$\\\hline
$u$&$\frac{1}{2}-\frac{4}{3}\sin^2\theta_W$&~$1/2$\\
$d$&$-\frac{1}{2}+\frac{2}{3}\sin^2\theta_W$&$-1/2$\\\hline
 \end{tabular}
 \caption{Couplings of the quarks~($u,d$) to $Z_\mu$ field.}\label{WS:gvga_quark}
 \end{center}
\end{table}

In analogy with the Lagrangian for the weak 
interaction for the leptons,  the most general Lagrangian for the weak charged current interaction, for the 4-quark flavor, is written as~\cite{Athar:2020kqn}:
\begin{eqnarray}\label{WS:CCquark}
 \mathcal{L}_{cc}^{int}(\text{quarks})&=&-\frac{g}{\sqrt{2}}\sum_q \left(\bar{q}_L \gamma^{\mu}\tau^+ q_L W_\mu^+ 
 +\bar{q}_L \gamma^{\mu}\tau^- q_L W_\mu^-\right) \\
 &=&-\frac{g}{2\sqrt{2}}\left[\left( \bar{u}\gamma^{\mu}(1-\gamma^{5})d^\prime +\bar{c}\gamma^\mu (1-\gamma^5)
 s^\prime  \right)W_\mu^+ +h.c. \right],~~~~~~~
\end{eqnarray}
where $d^\prime$ and $s^\prime$ are Cabibbo mixed states given by $d^\prime= d~cos\theta_C + s~sin\theta_C$ and $s^\prime= -d~ sin\theta_C + s~cos\theta_C$.

      \begin{figure}
 \begin{center}
     \includegraphics[height=8cm,width=10.5cm]{wq2_2gev.eps}
    \includegraphics[height=8cm,width=10.5cm]{kinematics_scattering_processes_3gev.eps}
       \end{center}
  \caption{Kinematical $Q^2-W$ plane depicting neutrino-nucleon scattering at two representative laboratory neutrino energies, i.e. 2~GeV~(upper panel) and 3~GeV~(lower panel). }
   \label{fig:kin}
 \end{figure}
 
Similarly, the neutral current weak interaction Lagrangian for the quarks is written  as~\cite{Athar:2020kqn}:
\begin{eqnarray}
 \mathcal{L}_{int}^{NC}&=&-j_{NC}^{\mu}(\text{quark})Z_{\mu},\label{WS:lagQ1} \\
 \text{with~~}\qquad j_{NC}^{\mu}(\text{quark})&=&\frac{e}{2 \sin\theta_W \cos\theta_W }\sum_q \bar{q}
 \gamma_{\mu}(g_{V}^{q}-g_{A}^{q}\gamma_5 )q ,\\
 \text{where~~}g_{V}^{q}&=&\frac{1}{2}\tau_{3}^{q}-2 \sin^2 \theta_W Q_q \text{~~~and }\label{WS:gv_quark}\\
 g_{A}^{q}&=&\frac{1}{2}\tau_{3}^{q}.\label{WS:ga_quark}
\end{eqnarray}
The values of $g_{V}$ and $g_{A}$ for $u$ and $d$ quarks are given in Table-\ref{WS:gvga_quark}.

However, certain phenomena, such as the neutrino mixing and oscillations--which imply that neutrinos possess small masses--as well as observed decay processes involving flavor-changing neutral currents~(FCNC) and lepton flavor violation~(LFV), remain unexplained by the Standard Model and suggest
the need for the physics Beyond the Standard Model~(BSM).

In this paper, we have studied the charged current induced (anti)neutrino scattering from the free nucleon target. In the (anti)neutrino energy region of a few GeV, relevant for most of the present and future neutrino oscillation experiments using the accelerator neutrinos like T2K, MicroBooNE, NOvA, DUNE, T2HK, etc., the neutrino cross section receives contribution from the quasielastic, inelastic and deep inelastic scattering processes. 
In general, when an (anti)neutrino interacts with a nucleon target, then the center of mass energy $W$, given by $W^2 = (p + q)^2$, where $q$ is the four-momentum transferred to the hadronic system, corresponds to the quasielastic scattering when $W = M$. As $W$ increases, the processes involving the single pion, multiple pion, kaon, eta and other meson production starts contributing and finally to the deep inelastic scattering~(DIS) region, where jet of hadrons are produced in the final state. The intermediate region, spanning the meson production threshold to the DIS domain, is referred to in the literature as the shallow inelastic scattering~(SIS) region. The kinematic region defining the SIS and the onset of DIS is not free from ambiguities, and to define these two kinematic regions the experiments use different cuts on $W$.  
To kinematically depict the different scattering processes relevant for the neutrino-nucleon cross sections in the few GeV energy region, in Fig.~\ref{fig:kin}, we have shown the $Q^2~(=-q^2 \ge 0)$ vs. $W$ plot at two different values of neutrino energies viz. 2 and 3~GeV. 
It may be observed from the figure that with the increase in $W$, several resonances contribute and besides pions, many other inelastic processes also start contributing to the neutrino-nucleon cross sections.

In Section~\ref{sec:formalism}, we discuss in brief the formalism of the aforementioned processes. 
In the case of inelastic scattering processes, we have studied the production of strange baryon like $\Lambda$, $\Sigma$, and mesons like pion, kaon, $\eta$ meson, in the final state as well as the associated particle production in which kaons and hyperons both are produced. 
The results for the $Q^2$ distribution of the cross section for these processes and their discussions are presented in Section~\ref{result}. Section~\ref{summary} concludes the findings of the present work.

\section{Formalism}\label{sec:formalism}
\subsection{Quasielastic scattering}
The general expression for the differential scattering cross section for the $\Delta S=0$ and $|\Delta S|=1$ quasielastic~(QE) reactions
\begin{eqnarray}
\label{eq:QE}
 \nu_{\mu}/\bar{\nu}_{\mu} (k)+ N(p) &\longrightarrow& \mu^{\mp} (k^{\prime}) + N^{\prime} (p^{\prime}); \qquad \quad N,N^{\prime} = n,p\\
 \label{eq:QE_Y}
 \bar{\nu}_{\mu} (k) + N(p) &\longrightarrow&  \mu^{+} (k^{\prime}) + Y (p^{\prime}); \qquad \quad Y=\Lambda,\Sigma^{0}, \Sigma^{-}
\end{eqnarray} 
is written as
\begin{eqnarray}\label{diff_xsect_quasi}
d\sigma=\frac{(2\pi)^{4}\delta^{4}(k+p-p^\prime-k^\prime)}{4(k\cdot p)}\frac {d{\vec{k}^{\;\prime}}}{(2\pi)^{3}2
E_{l}}\frac {d{\vec{p}^{\;\prime}}}{(2\pi)^{3}2E^{\prime}} {\bar\sum}\sum| {\cal M} |^2.
\end{eqnarray}
The quantities in the parentheses of Eqs.~(\ref{eq:QE}) and (\ref{eq:QE_Y}) are the four momenta of the corresponding particles
\Big($k=(E_\nu,\vec k),\;k^\prime=(E_l,\vec k^\prime),\;p^\prime=(E^\prime,\vec p^\prime),\;p=(E_p,\vec p)$\Big).

The invariant matrix element ${\cal M}$ is given by
\begin{eqnarray}\label{qe_lep_matrix}
{\cal M}=\frac{G_F}{\sqrt{2}}a~l_\mu~J^\mu,
\end{eqnarray}
where $G_F$ is the Fermi coupling constant, $a=\cos\theta_{C}~(\sin\theta_{C})$ for the $\Delta S=0~(1)$ processes, respectively, with $\theta_C$ being the Cabibbo angle. The leptonic weak current 
is given by
\begin{equation}\label{lep_curr}
l_\mu=\bar{u}(\vec{k}^{\prime})\gamma_\mu(1 \mp \gamma_5)u(\vec{k}),
\end{equation}
and $-(+)$ represents the neutrino~(antineutrino) induced QE scattering processes. The hadronic current~($J^\mu$) 
for CC induced interaction is given by
\begin{equation}\label{had_curr}
J^\mu=\bar{u}(\vec{p}^{\,\prime}) \;{\cal O}^{\mu}\; u(\vec{p}),
\end{equation}
where ${\cal O}^{\mu} = V^{\mu}-A^{\mu}$ is the weak hadronic vertex, and the matrix elements 
of the vector~($V_\mu$) and the axial-vector~($A_\mu$) currents are given by~\cite{Fatima:2018tzs}:
\begin{eqnarray}\label{vx}
\langle N^\prime(\vec{p}^{\,\prime}) | V^\mu| N(\vec{p}) \rangle &=& \bar{u}(\vec{p}^{\,\prime}) \left[ \gamma^\mu 
f_1^{NN^{\prime}(Y)}(Q^2)+i\sigma^{\mu \nu} \frac{q_\nu}{(M+M^\prime)} f_2^{NN^{\prime}(Y)}(Q^2) + \frac{2 q^\mu}{(M + M^\prime)} f_3^{NN^{\prime}(Y)}(Q^2) \right] u(\vec{p}),
\\
\label{vy}
\langle N^\prime(\vec{p}^{\,\prime}) | A^\mu| N(\vec{p}) \rangle &=& \bar{u} (\vec{p}^{\,\prime}) \left[ \gamma^\mu 
\gamma_5 g_1^{NN^{\prime}(Y)}(Q^2) +  i\sigma^{\mu \nu} \frac{q_\nu}{(M+M^\prime)} \gamma_5 g_2^{NN^{\prime}(Y)}(Q^2) +  \frac{2 q^\mu}{(M+M^\prime)} \gamma_5 
g_3^{NN^{\prime}(Y)}(Q^2) \right] u(\vec{p}).~~~~~~
\end{eqnarray}
In the above expression, $Q^2 = -q^2=-(k-k^{\prime})^2$ is the four momentum transfer squared, $M$ and 
$M^\prime$ are the masses of the initial and the final baryons, respectively. $f_1^{NN^{\prime}(Y)}(Q^2)$, $f_2^{NN^{\prime}(Y)}(Q^2)$ and $f_3^{NN^{\prime}(Y)} (Q^2)$ are 
the vector, induced weak magnetic and induced scalar $N-N^{\prime}$ or $N-Y$ transition form factors corresponding to the $\Delta S=0$ or $|\Delta S| = 1$ processes, respectively, and $g_1^{NN^{\prime}(Y)}(Q^2)$, $g_2^{NN^{\prime}(Y)}(Q^2)$ and $g_3^{NN^{\prime}(Y)}(Q^2)$ are the axial-vector, 
induced tensor~(or weak electric) and induced pseudoscalar form factors, respectively. 
These form factors have been determined using the various symmetry properties of the weak hadronic currents like the time-reversal~(T) invariance, G-parity~(G) invariance, conserved axial-vector current~(CVC) and partially conserved axial-vector current~(PCAC) hypotheses, etc., and the details  can be found in Refs.~\cite{SajjadAthar:2022pjt, Fatima:2018tzs}. 
The time reversal invariance implies that all the vector and axial vector form factors are real. G-parity is a multiplicative quantum number that results from the generalization of C-parity to multiplets of mesons~(like pions, rho, etc.), and G-parity$~(G=Ce^{i\pi I_y})$ is defined as a combination of charge conjugation~(C) and a 180$^{\circ}$ rotation around the y-axis of isospin space ($I_y)$. 
The concept of G-invariance was first introduced by Weinberg to classify the above mentioned form factors as first and second class current form factors. 
The first class vector and axial vector form factors are those form factors that transform in the same way as the basic form factors $f_{1}(Q^2)$ and $g_{1}(Q^2)$ do, while the second class current form factors are those that transform with an opposite sign in their respective currents. 
With this definition the form factors $f_{1}(Q^2)$, $f_{2}(Q^2)$, $g_{1}(Q^2)$, and $g_{3}(Q^2)$ are classified as first class currents while $f_{3}(Q^2)$ and $g_{2}(Q^2)$ are classified as the second class current form factors. 
If G invariance is assumed to be an exact symmetry, then the form factors associated with the second class currents vanish. 
In the vector sector, the form factor $f_{3}(Q^2)$ is further constrained to vanish due to CVC. Therefore, to study the effect of G-invariance one takes $g_{2}(Q^2)$ to be non-zero in the numerical calculations.  
Moreover, if one assumes $g_{2}(Q^2)$ to be purely real, then it preserves T invariance, while the complex or purely imaginary values of $g_{2}(Q^2)$ results in T violation in the weak sector along with G violation.

 The $Q^2$ distribution for the cross section, i.e., $d\sigma/dQ^2$ for the processes given in 
 Eqs.~(\ref{eq:QE}) and (\ref{eq:QE_Y}) is written as
\begin{equation}\label{dsig}
 \frac{d\sigma}{dQ^2}=\frac{G_F^2 a^2}{8 \pi {M}^2 {E^2_\nu}} N(Q^2),
\end{equation}
where $N(Q^2) = \cal{J}^{\mu \nu} \cal{L}_{\mu \nu}$ and the expression of $N(Q^2)$ is given in the Appendix-I of Ref.~\cite{Fatima:2018tzs}.

\subsection{Inelastic scattering}
\begin{table}
\begin{center}
		%     \vspace{1cm}
\begin{tabular*}{160mm}{@{\extracolsep{\fill}}c c c}
			%       \noalign{\vspace{-8pt}}
	\hline \hline
	~~~Process               & CC induced neutrino reactions \qquad & CC induced antineutrino reactions~~~\\ \hline
 & $\nu_{\mu}  + p \longrightarrow \mu^{-} + p + \pi^{+}$ & $\bar{\nu}_{\mu} + p \longrightarrow \mu^{+} + n+ \pi^{0}$ \\ 
			
Single pion production & $\nu_{\mu} + n \longrightarrow \mu^{-} + p + \pi^0$ & $\bar{\nu}_{\mu} + p \longrightarrow \mu^{+} + p + \pi^-$ \\ 
			
 & $\nu_{\mu} + n \longrightarrow \mu^{-} + n + \pi^+ $ &  $\bar{\nu}_{\mu} + n \longrightarrow \mu^{+} + n + \pi^-$\\ 
			
Eta production & $\nu_{\mu} + n \longrightarrow \mu^{-} + p + \eta$ & $\bar{\nu}_{\mu} + n \longrightarrow \mu^{+} + n + \eta$ \\ 
			
Associated particle production & $\nu_{\mu} + n \longrightarrow \mu^{-} + \Lambda + K^{+}$ &  $\bar{\nu}_{\mu} + p \longrightarrow \mu^{+} + \Lambda + K^{0}$\\ 

& $\nu_{\mu} + p \longrightarrow \mu^{-} + p + K^{+}$ & $\bar{\nu}_{\mu} + p \longrightarrow \mu^{+} + p + K^{-}$\\ 
Single Kaon production& $\nu_{\mu} + n \longrightarrow \mu^{-} + p + K^{0}$ & $\bar{\nu}_{\mu} + p \longrightarrow \mu^{+} + n + \bar{K}^{0}$\\
& $\nu_{\mu} + n \longrightarrow \mu^{-} + n + K^{+}$ & $\bar{\nu}_{\mu} + n \longrightarrow \mu^{+} + n + K^{-}$\\
\hline \hline
\end{tabular*}
\end{center}
		%%changehere
\caption{CC neutrino and antineutrino induced IE processes considered in the present work. }\label{sec2:Table1}
		%    \vspace{15mm}
\end{table}
The general expression for the differential scattering cross section of the inelastic process~(Table-\ref{sec2:Table1})
\begin{equation}\label{eq:inelastic:reaction}
\nu_{l}/\bar{\nu}_{l} (k) + N(p) \longrightarrow l^{\mp} (k^{\prime}) + B(p^{\prime}) + m(p_{m})
\end{equation}
\begin{figure}
 \includegraphics[height=5 cm, width=0.9\textwidth]{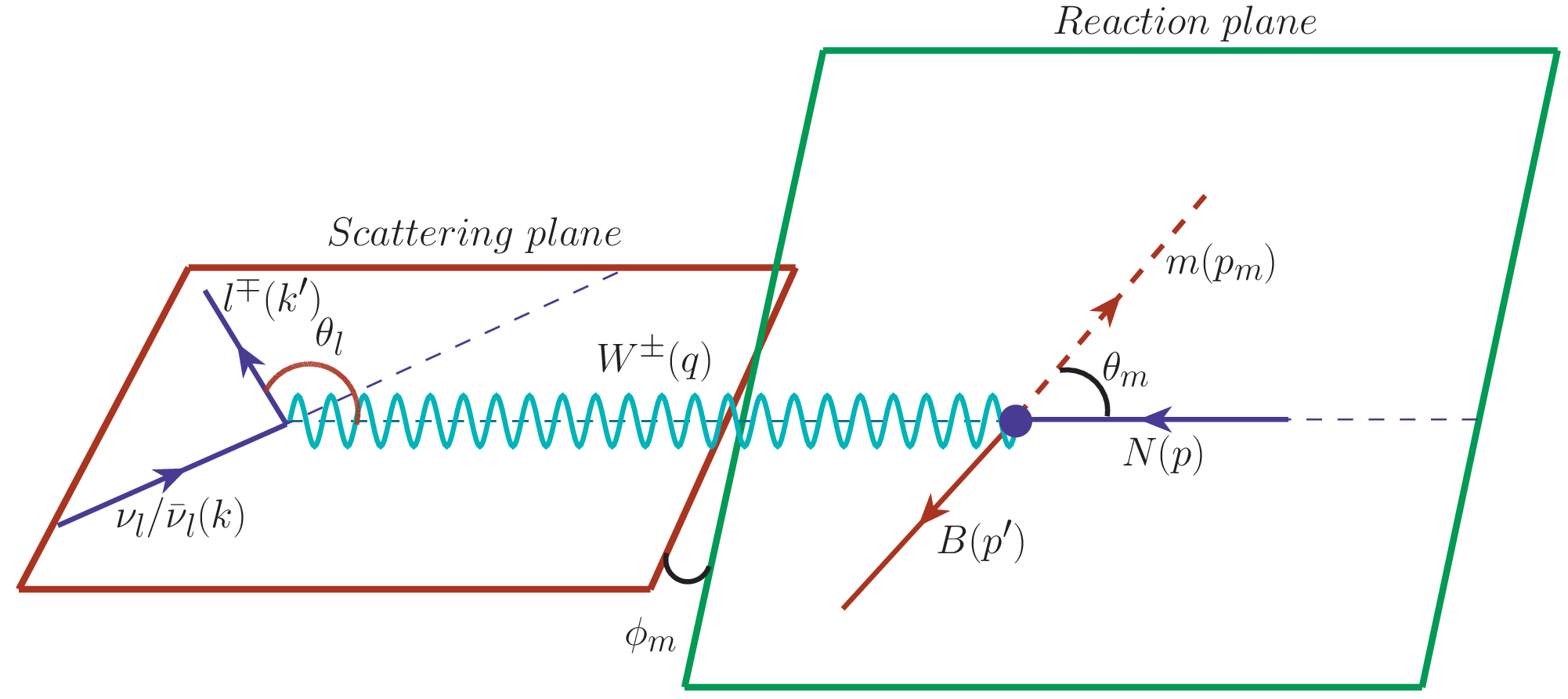}
 \caption{(Anti)neutrino scattering and reaction planes.}
 \label{reactionplane}
\end{figure}
in the laboratory frame is given by
\begin{eqnarray}\label{eq:sigma_inelas}
d\sigma &=& \frac{1}{4 ME_\nu(2\pi)^{5}} \frac{d{\vec k}^{\prime}}{ (2 E_{l})} 
\frac{d{\vec p\,}^{\prime}}{(2 E_{B})} \frac{d{\vec p}_{m}}{ (2 E_m)}
 \delta^{4}(k+p-k^{\prime}-p^{\prime}-p_{m})\overline{\sum}\sum | \mathcal M |^2,\;\;\;\;\;
\end{eqnarray}
where in Eq.~(\ref{eq:inelastic:reaction}), $m~(=\pi, \eta, K$, etc.) is a meson produced with a baryon~($B=N,Y$, etc.) in 
the final state. $ k( k^\prime) $ is the four momentum of the incoming~(outgoing) lepton having energy $E_\nu( E_l)$; $p$ is 
the four momentum of the incoming nucleon which is at rest, $E_B$ and $p^\prime$ are respectively the energy and four 
momentum of the outgoing baryon, and the meson four momentum is $p_m$ with energy $ E_m$, and $M$ is the nucleon mass. 
$\overline{\sum}\sum | \mathcal M |^2  $ is the square of the transition amplitude averaged~(summed) over the 
spins of the initial~(final) states and the transition matrix element is written in terms of the leptonic and the hadronic 
currents as 
\begin{equation}
\label{eq:Gg}
 \mathcal M = a \frac{G_F}{\sqrt{2}}\, {l_\mu} J^{\mu},
\end{equation}
where the leptonic current $l_\mu$, and the constants $a$, 
$\theta_C$ and $G_F$ are defined after Eq.~(\ref{qe_lep_matrix}). $  J^{\mu}$  is the hadronic current for $W^{\pm} + N 
\longrightarrow B + m$  interaction for CC 
induced processes.

The differential scattering cross section $\frac{d\sigma}{dQ^2}$, for the reactions shown in 
Eq.~(\ref{eq:inelastic:reaction}) is expressed as
\begin{equation}\label{sigma:weak}
 \frac{d\sigma}{dQ^2} = \int_{W_{min}}^{W_{max}} dW \int_{0}^{2\pi} d\phi_{qp_{m}} \int_{E_{m}^{min}}^{E_{m}^{max}} dE_{m} \frac{1}{(2\pi)^{4}} \frac{1}{64E_{\nu}^{2}M^2} \frac{W}{|\vec{q}\;|} \overline{\sum} \sum |{\cal M}|^2, 
\end{equation}
where $E_{m}^{min}$ and $E_{m}^{max}$ are respectively the minimum and maximum energy of the outgoing meson and $\phi_{qp_{m}}$ is the azimuthal angle between the scattering plane and the reaction plane as shown in Fig.\ref{reactionplane}.

\begin{figure}
\begin{center}
 \includegraphics[height=3.5 cm, width=0.9\textwidth]{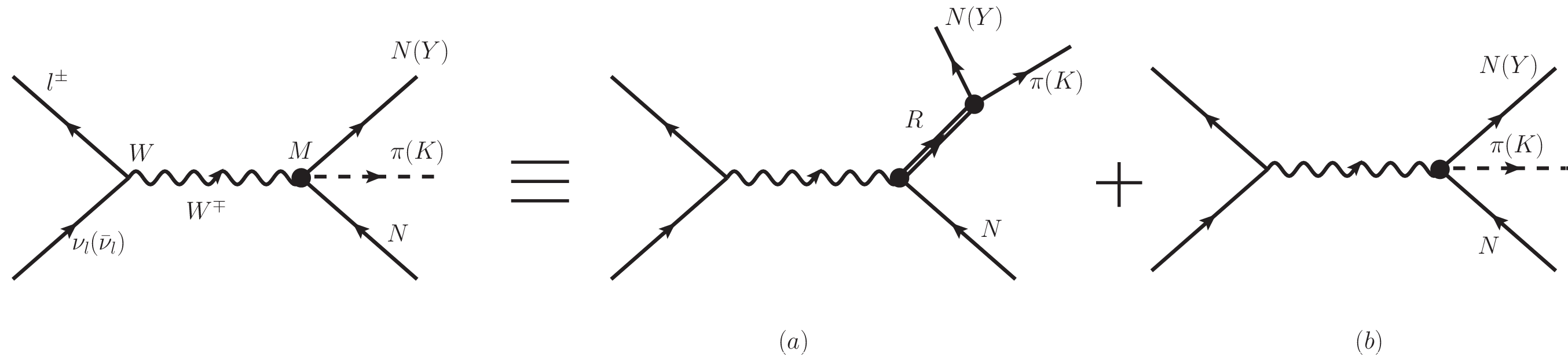}
 \end{center}
 \caption{Feynman diagram for CC inelastic scattering processes.}
 \label{Fdiagram1}
\end{figure}

\begin{figure}
\begin{center}
 \includegraphics[height=3.5 cm, width=0.9\textwidth]{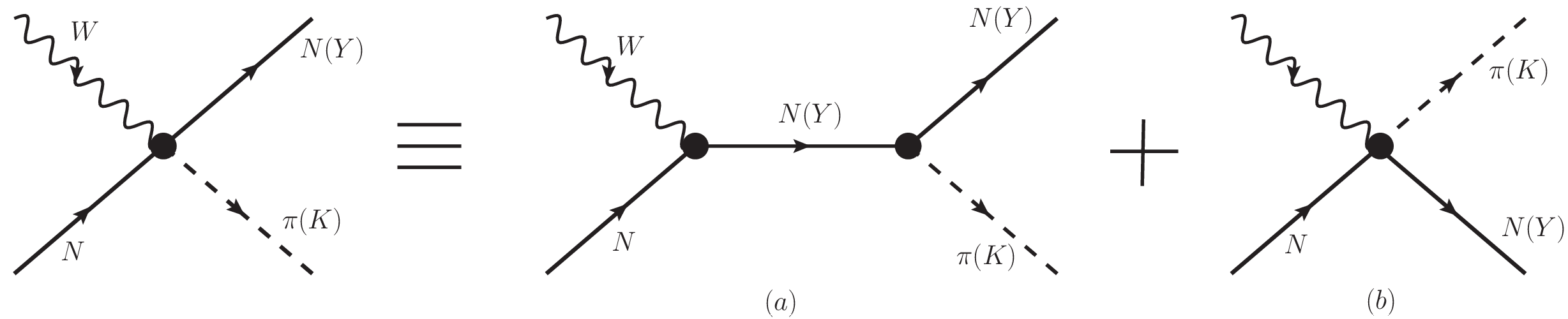}
 \end{center}
 \caption{Feynman diagram for nonresonant background terms contributing to the inelastic processes.}
  \label{Fdiagram2}
\end{figure}

    \begin{table}
  \begin{center}
    \begin{tabular*}{170mm}{@{\extracolsep{\fill}}c c c c  c c cc c}
      \hline \hline
  Resonance           & $M_R$ & $\Gamma$ &  $I(J^P) $    &  \multicolumn{5}{c}{Branching Ratios~(in $\%$)}  \\
   & (GeV) & (GeV)&  & $N\pi$& $N\eta$ & $K\Lambda$ & $K\Sigma$&$\pi\pi N$ \\ \hline
      
  $P_{11} (1440)$  & $1.370 \pm 0.01$ & $0.175 \pm 0.015$ & $1/2(1/2^+)$ & $65$ & $<1$ & - & -&34 \\ \hline
      
  $S_{11} (1535)$  & $1.510 \pm 0.01$ & $0.130 \pm 0.020$ & $1/2(1/2^-)$ & $42$ & $42$ & - & -& 8\\ \hline
      
  $S_{31} (1620)$  & $1.600 \pm 0.01$ & $0.120 \pm 0.020$ & $3/2(1/2^-)$ & $30$ & - & - & -& 67\\ \hline
   
  $S_{11} (1650)$  & $1.655 \pm 0.015$ & $0.135 \pm 0.035$ & $1/2(1/2^+)$ & $60$ & $25$ & $10$ & -&5 \\ \hline
     
  $P_{11} (1710)$  & $1.700 \pm 0.02$ & $0.120 \pm 0.040$ & $1/2(1/2^+)$ & $10$ & $30$ & $15$ & $<1$&- \\ \hline
      
  $P_{11} (1880)$  & $1.860 \pm 0.04$ & $0.230 \pm 0.050$ & $1/2(1/2^+)$ & $6$ & $30$ & $20$ & $17$&44 \\ \hline
      
  $S_{11} (1895)$  & $1.910 \pm 0.02$ & $0.110 \pm 0.030$ & $1/2(1/2^-)$ & $10$ & $25$ & $18$ & $13$&- \\ \hline
%      
%  $S_{31} (1900)$  & $1.865 \pm 0.035$ & $0.240 \pm 0.060$ & $3/2(1/2^-)$ & $8$ & - & - & -& \\ \hline
      
      $P_{33} (1232)$  & $1.210 \pm 0.001$ & $0.100 \pm 0.002$ & $3/2(3/2^+)$  & $99.4$ & - & - & -&- \\ \hline
      
 $D_{13} (1520)$  & $1.510 \pm 0.005$ & $0.110 \pm^{0.010}_{0.005}$ & $1/2(3/2^-)$  & $60$ & - & - & - & 30
 \\ \hline
%      
% $P_{33} (1600)$  & $1.510 \pm 0.05$ & $0.270 \pm 0.07$ &  $3/2(3/2^+)$  & $16$ & - & - & - \\ \hline
      
% $D_{13} (1700)$  & $1.700 \pm 0.05$ & $0.200 \pm 0.100$ & $1/2(3/2^-)$ & $12$ & $12$ & - & - &\\ \hline
      
 $D_{33} (1700)$  & $1.665 \pm 0.025$ & $0.250 \pm 0.05$ & $3/2(3/2^-)$ & $15$ & - & - & - &32\\ \hline
      
 $P_{13} (1720)$  & $1.675 \pm 0.015$ & $0.250 \pm_{0.150}^{0.100}$ & $1/2(3/2^+)$& $11$ & $3$ & $4.5$ & - &70
 \\ \hline
      
% $D_{13} (1875)$  & $1.900 \pm 0.05$ & $0.160 \pm 0.06$ & $1/2(3/2^-)$ & $7$ & $<1$ & $0.2$ & $0.7$& \\ \hline
      
 $P_{13} (1900)$  & $1.920 \pm 0.02$ & $0.150 \pm 0.05$ & $1/2(3/2^+)$ & $10$ & $8$ & $11$ & $5$&60 \\ \hline \hline
%  $P_{31} (1910)$  & $1.860 \pm 0.03$ & $0.300 \pm 0.100$ & $3/2$ & $+$ & $20\%$ & - & - & $9\%$ \\ \hline
%      
%  $S_{31} (2150)$  & $2.140 \pm 0.08$ & $0.200 \pm 0.080$ & $3/2$ & $-$ & $8\%$ & - & - & - \\ \hline  \hline
  \end{tabular*}
  \end{center}
\caption{Properties of the spin $1/2$ and $3/2$ resonances available in the PDG~\cite{ParticleDataGroup:2020ssz}, with 
Breit-Wigner mass $M_R$, the total decay width $\Gamma$, isospin $I$, spin $J$, parity $P$, and the central value of the 
branching ratio into different meson-baryon like $N\pi$, $N\eta$, $K \Lambda$, $K \Sigma$ and $\pi \pi N$.}
\label{Tab:Resonance}
%    \vspace{15mm}
\end{table}

The hadronic 
current~($j^{\mu}$) receives contribution from the nonresonant background~(NRB) terms as well as from the resonance excitations as shown in Fig.\ref{Fdiagram1} 
and their decay into a particular meson-baryon final state. The different inelastic channels receive contribution from the 
different background terms~(Fig.\ref{Fdiagram2}) as well as from the different resonance excitations. 
Table~\ref{Tab:Resonance} lists the resonances considered in the present work along with their properties  like the mass, decay width, spin, isospin, parity, and the branching ratios of the resonances into the different meson-baryon channels.
Specifically, the hadronic current for an inelastic scattering process is written as
\begin{equation}\label{allterms}
 J^{\mu}=J^{\mu}_{\text NR} ~+~ J^{\mu}_{R_{\frac{1}{2}}}~+~J^{\mu}_{R_{\frac{3}{2}}}, 
\end{equation}
where $J^{\mu}_{\text NR}$, $J^{\mu}_{R_{\frac{1}{2}}}$, and $J^{\mu}_{R_{\frac{3}{2}}}$, respectively, represent the 
contribution of the hadronic current from the NRB terms, spin $\frac{1}{2}$ resonance, and spin 
$\frac{3}{2}$ resonance excitations. 
The 
structure of the hadronic currents for the background and the resonance terms for the different inelastic channels has been discussed in detail in Ref.~\cite{SajjadAthar:2022pjt}.

\subsection{Deep inelastic scattering~(DIS)}
The general expression of the double differential scattering cross section for CC induced $\nu_l 
({\bar\nu}_l)-N$ DIS in the laboratory frame corresponding to the reaction:
\begin{eqnarray}\label{reaction}
\nu_l(k) / \bar\nu_l(k) + N(p) \rightarrow l^-(k') / l^+(k') + X(p');~~~~~~~~~~~ l=e, \mu, \tau,
\end{eqnarray}
  is given by~\cite{Ansari:2020xne, Athar:2020kqn}:
  \begin{equation}
\label{eq:q2nu}
\frac{ d^2\sigma  }{ dx dy } =  \frac{y M}{\pi }~\frac{E_\nu}{E_l}~\frac{|{\vec k^\prime}|}{|{ \vec k}|}\;   \overline\sum \sum |{\cal M}|^2, 
\end{equation}
%  \begin{equation}
% \label{eq:q2nu}
% { d^2\sigma_N^{WI} \over d\Omega' dE'} =  \frac{1}{2\pi^2 }~\frac{|{\vec{k}^\prime}|}{|{ \vec{k}}|}\;   \overline{\sum} \sum 
% |{\cal M}|^2 \;.
% \end{equation}
%\cite{Ansari:2020xne}
 where the four momentum transfer $q=k - k^\prime$, the Bjorken variable $x= \frac{Q^2}{2p\cdot q}$ and the inelasticity $y= \frac{p\cdot q}{p \cdot k}$ with $Q^2(\ge 0)= -q^2$. 
 
The matrix element square i.e. $\overline{\sum} \sum |{\cal M}|^2$ in Eq.~(\ref{eq:q2nu}), 
averaged over the initial spin states and summed over the final spin states, is given in terms of the leptonic~($L_{\mu\nu}$) 
and hadronic~($ W^{\mu\nu}_N$) tensors as:
\begin{equation}\label{amp_wk}
\overline{\sum} \sum |{\cal M}|^2 = \frac{G_F^2}{2}~\left(\frac{M_W^2}{Q^2+M_W^2}\right)^2 ~L_{\mu\nu} ~W^{\mu\nu}_N,
\end{equation}
where $M_W$ is the mass of the intermediate vector boson $W^\pm$. 
Using Eq.~(\ref{lep_curr}), the leptonic tensor is given by
\begin{eqnarray}\label{lep_weak}
L_{\mu \nu} &=&\overline\sum\sum l_\mu l_\nu^\dagger=8(k_{\mu}k'_{\nu}+k_{\nu}k'_{\mu}
-k.k^\prime g_{\mu \nu}  \pm i \epsilon_{\mu \nu \rho \sigma} k^{\rho} 
k'^{\sigma})\,.
\end{eqnarray}
In the case of DIS, the hadronic final state is not observed, 
therefore, the hadronic tensor $W^{\mu\nu}_N$ is written to parameterize our ignorance of the hadronic current. The most 
general form of the hadronic tensor is constructed by using the available four vectors at the disposal of hadronic vertex, 
i.e. the four momentum $p^{\mu}$, the four momentum transfer $q^\mu$, and the metric tensor $g^{\mu\nu}$ and is given by:
\begin{eqnarray}\label{ch2:had_ten_N}
W_{N}^{\mu \nu} &=&
\left( \frac{q^{\mu} q^{\nu}}{q^2} - g^{\mu \nu} \right) \;
W_{1N} (\nu, Q^2)
+ \frac{W_{2N} (\nu, Q^2)}{M^2}\left( p^{\mu} - \frac{p . q}{q^2} \; q^{\mu} \right)
 \nonumber\\
&\times&\left( p^{\nu} - \frac{p . q}{q^2} \; q^{\nu} \right)-\frac{i}{2M^2} \epsilon^{\mu \nu \rho \sigma} p_{ \rho} q_{\sigma}
W_{3N} (\nu, Q^2) + \frac{W_{4N} (\nu, Q^2)}{M^2} q^{\mu} q^{\nu}\nonumber\\
&&+\frac{W_{5N} (\nu, Q^2)}{M^2} (p^{\mu} q^{\nu} + q^{\mu} p^{\nu})
+ \frac{i}{M^2} (p^{\mu} q^{\nu} - q^{\mu} p^{\nu})
W_{6N} (\nu, Q^2)\,,
\end{eqnarray}
where $\nu=E_\nu-E_l$ is the energy transfer.
The contribution of the term with $W_{6N} (\nu, Q^2)$ vanishes when contracted with the leptonic tensor. In the limit $Q^2\to\infty$ and $\nu\to\infty$,
the structure functions $W_{iN}  (\nu,Q^2);~(i=1-5)$ are generally written in terms of the dimensionless nucleon 
structure functions $F_{iN}  (x),\;\;i=1 - 5$ as~\cite{Ansari:2020xne}: 
 \begin{eqnarray}\label{ch2:relation}
 F_{1N}(x) &=& W_{1N}(\nu,Q^2), \;\;\;\;
 F_{2N}(x) = \frac{Q^2}{2xM^2}W_{2N}(\nu,Q^2),\;\;\;
 F_{3N}(x) = \frac{Q^2}{xM^2}W_{3N}(\nu,Q^2), \nonumber\\
 F_{4N}(x) &=& \frac{Q^2}{2M^2}W_{4N}(\nu,Q^2), \;\;\;
 F_{5N}(x) = \frac{Q^2}{2xM^2}W_{5N}(\nu,Q^2) \nonumber
\end{eqnarray}
The simplified expression of hadronic tensor in terms of the dimensionless nucleon structure functions $F_{iN}(x,Q^2)\;\;(i=1-5)$ is given by:
\begin{eqnarray}\label{had_weak_red}
W_{N}^{\mu \nu} 
& = &- g_{\mu\nu} F_{1N}(x,Q^2) + \frac{p_{\mu}p_{\nu}}{{p_\cdot q}} F_{2N}(x,Q^2) - i \epsilon_{\mu\nu\rho\sigma} \frac{p^{\rho} q^{\sigma}}{2 p_1\cdot q} F_{3N}(x,Q^2) + \nonumber \\
          &   & \frac{q_{\mu} q_{\nu}}{p\cdot q} F_{4N}(x,Q^2) +
                (p_{\mu}q_{\nu} + p_{\nu}q_{\mu}) F_{5N}(x,Q^2).
\end{eqnarray}
The expression for the differential scattering cross section for the $\nu_l/{\bar\nu}_l - N$ scattering 
given in Eq.~\ref{eq:q2nu} is written by using Eqs.~\ref{lep_weak} and \ref{had_weak_red} as~\cite{Ansari:2020xne}:
\begin{eqnarray}
 \frac{d^2\sigma}{dxdy}&=&\frac{G_F^2ME_\nu}{\pi(1+\frac{Q^2}{M_W^2})^2}
 \Big\{\Big[y^2x+\frac{m_l^2 y}{2E_\nu M}\Big]F_{1N}(x,Q^2)+
 \Big[\Big(1-\frac{m_l^2}{4E_\nu^2}\Big)-\Big(1+\frac{Mx}{2E_\nu}\Big)y\Big]F_{2N}(x,Q^2)\nonumber\\
 &\pm& \Big[xy\Big(1-\frac{y}{2}\Big)-
 \frac{m_l^2 y}{4E_\nu M}\Big]F_{3N}(x,Q^2)
 +\frac{m_l^2(m_l^2+Q^2)}{4E_\nu^2M^2 x}F_{4N}(x,Q^2)-\frac{m_l^2}{E_\nu M}F_{5N}(x,Q^2)\Big\},\;\;\;
\end{eqnarray}
where $m_l$ is the mass of the final state charged lepton. $x$ and $y$ are the 
scaling variables which lie in the following ranges:
\begin{eqnarray}
\frac{m_l^2}{2M (E_\nu - m_l)} \le x \le 1;&~~~&a-b \le y\le a+b, \qquad \qquad \text{with} \\
 a=\frac{1-m_l^2\Big(\frac{1}{2ME_\nu x}+\frac{1}{2E_\nu^2} \Big)}{2\Big(1+\frac{M x}{2E_\nu}\Big)},&\;\;&
 b=\frac{\sqrt{\left(1-\frac{m_l^2}{2 M E_\nu x}\right)^2-\frac{m_l^2}{E_\nu^2}}}{2\Big(1+\frac{M x}{2E_\nu}\Big)}.
\end{eqnarray}
However, for the massless lepton limit~($m_l\to0$; $l=e,\mu$), only the first three structure functions contribute to the scattering cross section and the 
above expression reduces to:
\begin{eqnarray}
 \frac{d^2\sigma}{dxdy}&=&\frac{G_F^2ME_\nu}{\pi(1+\frac{Q^2}{M_W^2})^2}
 \Big\{\Big[y^2x+\frac{m_l^2 y}{2E_\nu M}\Big]F_{1N}(x,Q^2)+
 \Big[\Big(1-\frac{m_l^2}{4E_\nu^2}\Big)-\Big(1+\frac{Mx}{2E_\nu}\Big)y\Big]F_{2N}(x,Q^2)\nonumber\\
 &\pm& \Big[xy\Big(1-\frac{y}{2}\Big)-
 \frac{m_l^2 y}{4E_\nu M}\Big]F_{3N}(x,Q^2)
 \Big\}.\;\;\;
\end{eqnarray}

  \begin{figure}
 \begin{center}
    \includegraphics[height=8cm,width=7.5cm]{dsigma_dQ2_neutrino_MA.eps}
    \hspace{3mm}
     \includegraphics[height=8cm,width=7.5cm]{dsigma_dQ2_antineutrino_MA.eps} 
       \end{center}
  \caption{$\frac{d\sigma}{dQ^2}$ vs. $Q^2$ for the processes $\nu_{\mu} + n \longrightarrow \mu^{-} + p$~(left panel) and $\bar{\nu}_{\mu} + p \longrightarrow \mu^{+} + n$~(right panel) at $E_{\nu}=2$~GeV at different values of $M_{A}$ viz. $M_{A} = 1.026$~GeV~(solid line), 1.1~GeV~(dashed line), 1.2~GeV~(dashed-dotted line), and 1.3~GeV~(double-dotted-dashed line).}
   \label{fig:QE_MA}
 \end{figure}

In general, the dimensionless nucleon structure functions are derived in the quark-parton model assuming Bjorken scaling and 
are functions of only one variable $x$. In this model, these structure functions obey Callan-Gross~\cite{Callan:1969uq} and 
Albright-Jarlskog~\cite{Albright:1974ts} relations, respectively, given by
\begin{eqnarray}
 F_{1}(x)&=&\frac{F_2(x)}{2 x}\;;\;\;
 F_{5}(x)=\frac{F_2(x)}{2 x}\;;\;\;F_4(x)=0. \nonumber
\end{eqnarray}
At the leading order of perturbative QCD, the structure functions are derived in terms of the parton distribution functions 
$q_i(x)$ and $\bar q_i(x)$ as:
\begin{eqnarray}\label{parton_wk}
F_{2}(x)  &=& \sum_{i} x [q_i(x) +\bar q_i(x)] \;;\;~~~~~~~~
x F_3(x) =  \sum_i x [q_i(x) -\bar q_i(x)].
\end{eqnarray} 
 For the present numerical calculations, we have used the MMHT nucleonic parton distribution functions~(PDFs) parameterization~\cite{Harland-Lang:2014zoa} in the MSbar scheme. 

\begin{figure}
 \begin{center}
    \includegraphics[height=8cm,width=7.5cm]{dsigma_dQ2_neutrino_g2R.eps}
    \hspace{3mm}
    \includegraphics[height=8cm,width=7.5cm]{dsigma_dQ2_antineutrino_g2R.eps} 
       \end{center}
  \caption{$\frac{d\sigma}{dQ^2}$ vs. $Q^2$ for the processes $\nu_{\mu} + n \longrightarrow \mu^{-} + p$~(left panel) and $\bar{\nu}_{\mu} + p \longrightarrow \mu^{+} + n$~(right panel) at $E_{\nu}=2$~GeV at the different values of $g_{2}^{R} (0)$ viz. $g_{2}^{R} (0) = 0$~(solid line),  +3~(double-dotted-dashed line),  and $-3$~(solid line with triangle).}
   \label{fig:QE_g2R}
 \end{figure}
 
\section{Results and discussions}
\label{result}
\subsection{Quasielastic scattering}
 \begin{figure}
 \begin{center}
    \includegraphics[height=8cm,width=7.5cm]{dsigma_dQ2_neutrino_g2I.eps}
    \hspace{3mm}
     \includegraphics[height=8cm,width=7.5cm]{dsigma_dQ2_antineutrino_g2I.eps} 
       \end{center}
  \caption{$\frac{d\sigma}{dQ^2}$ vs. $Q^2$ for the processes $\nu_{\mu} + n \longrightarrow \mu^{-} + p$~(left panel) and $\bar{\nu}_{\mu} + p \longrightarrow \mu^{+} + n$~(right panel) at $E_{\nu}=2$~GeV at two different values of $g_{2}^{I} (0)$ viz. $|g_{2}^{I} (0)| = 0$~(solid line) and 3~(double-dotted-dashed line).}
   \label{fig:QE_g2I}
 \end{figure}

 \begin{figure}
 \begin{center}
    \includegraphics[height=8cm,width=7.5cm]{dsigma_dq2_Lambda_g2R.eps}
    \hspace{3mm}
     \includegraphics[height=8cm,width=7.5cm]{dsigma_dq2_SigmaM_g2R.eps} 
       \end{center}
  \caption{$\frac{d\sigma}{dQ^2}$ vs. $Q^2$ for the processes $\bar{\nu}_{\mu} + p \longrightarrow \mu^{+} + \Lambda$~(left panel) and $\bar{\nu}_{\mu} + n \longrightarrow \mu^{+} + \Sigma^{-}$~(right panel) at $E_{\bar{\nu}_{\mu}}=2$~GeV at different values of $g_{2}^{R} (0)$. 
  Lines and points have the same meaning as in Fig.~\ref{fig:QE_g2R}.}
   \label{fig:QE_Y_g2R}
 \end{figure}
 
 \begin{figure}
 \begin{center}
    \includegraphics[height=8cm,width=7.5cm]{dsigma_dq2_Lambda_g2I.eps}
    \hspace{3mm}
      \includegraphics[height=8cm,width=7.5cm]{dsigma_dq2_SigmaM_g2I.eps} 
       \end{center}
  \caption{$\frac{d\sigma}{dQ^2}$ vs. $Q^2$ for the processes $\bar{\nu}_{\mu} + p \longrightarrow \mu^{+} + \Lambda$~(left panel) and $\bar{\nu}_{\mu} + n \longrightarrow \mu^{+} + \Sigma^{-}$~(right panel) at $E_{\nu}=2$~GeV at different values of $g_{2}^{I} (0)$. 
  Lines and points have the same meaning as in Fig.~\ref{fig:QE_g2I}.}
   \label{fig:QE_Y_g2I}
 \end{figure}

\begin{figure}  
 \begin{center}
    \includegraphics[height=7cm,width=8cm]{xsec_comp2.eps} 
    \end{center}
\caption{$\sigma$ vs. $E_{\bar{\nu}_\mu}$ for the $\Lambda$ production. Solid~(dashed) line represents the result using $M_A = 
1.026~(1.2)$ GeV. Experimental results:  triangle 
right~\cite{Eichten:1972bb}, triangle up~\cite{Erriquez:1977tr}, square~\cite{Erriquez:1978pg}, triangle down~($\sigma = 
2.6^{+5.9}_{-2.1} \times 10^{-40} cm^2$)~\cite{Fanourakis:1980si}, circle~\cite{SKAT:1989nel} are shown with error bars. Theoretical curves are of 
Kuzmin and Naumov~\cite{Kuzmin:2008zz}~(double dashed-dotted line), Brunner et al.~\cite{SKAT:1989nel}~(dashed line), Erriquez 
et al.~\cite{Erriquez:1978pg}~(dashed-double dotted line) obtained using Cabibbo theory with axial-vector dipole mass as 
1~GeV, 1.1 GeV and 1 GeV, respectively, while the results of Wu et al.~\cite{Wu:2013kla}~(dotted line) and Finjord and 
Ravndal~\cite{Finjord:1975zy}~(dashed dotted line) are obtained using quark model.}\label{free_hyperon}
 \end{figure}
 
    \begin{figure}
 \begin{center}
    \includegraphics[height=7cm,width=7.5cm]{dsigma_dq2_p_pi+_nu.eps}
    \hspace{3mm}
      \includegraphics[height=7cm,width=7.5cm]{dsigma_dq2_n_pi-_antinu.eps} 
\\
    \includegraphics[height=7cm,width=7.5cm]{dsigma_dq2_n_pi+_nu.eps}
    \hspace{3mm}
      \includegraphics[height=7cm,width=7.5cm]{dsigma_dq2_p_pi-_antinu.eps} 
\\ 
\includegraphics[height=7cm,width=7.5cm]{dsigma_dq2_p_pi0_nu.eps}
    \hspace{3mm}
      \includegraphics[height=7cm,width=7.5cm]{dsigma_dq2_n_pi0_antinu.eps} 
       \end{center}
  \caption{$\frac{d\sigma}{dQ^2}$ vs. $Q^2$ for the processes $\nu_{\mu} + p \longrightarrow \mu^{-} + p + \pi^{+}$~(upper left panel), $\bar{\nu}_{\mu} + n \longrightarrow \mu^{+} + n + \pi^{-}$~(upper right panel), $\nu_{\mu} + n \longrightarrow \mu^{-} + n + \pi^{+}$~(middle left panel), $\bar{\nu}_{\mu} + p \longrightarrow \mu^{+} + p + \pi^{-}$~(middle right panel), $\nu_{\mu} + n \longrightarrow \mu^{-} + p + \pi^{0}$~(lower left panel) and $\bar{\nu}_{\mu} + p \longrightarrow \mu^{+} + n + \pi^{0}$~(lower right panel) at $E_{\nu}=2$~GeV. 
  Solid~(dashed-dotted) line represents the results obtained using the full model, where the full model receives contribution from the different non-resonant background terms as well as from the spin $\frac{1}{2}$ and $\frac{3}{2}$ resonances as discussed in the text,  when a cut on $W<2$~GeV~($W<1.4$~GeV) is applied. 
  Dashed~(dotted) line represents the results obtained using the background and $P_{33}(1232)$ resonance, when a cut on $W<2$~GeV~($W<1.4$~GeV) is applied.}
   \label{fig:pi}
 \end{figure}
 
 The numerical results for the differential and total cross sections are obtained using the BBBA05~\cite{Bradford:2006yz} parameterization of the Sachs' electric and magnetic form factors for the isovector vector form factors $f_{1,2}(Q^2)$, axial vector form factor having dipole parameterization i.e. $g_1(Q^2)= \frac{g_1(0)}{\left(1 + \frac{Q^2}{M_A^2}\right)^2}$, $g_1(0)~(=1.26)$ is the axial charge and  $M_A$ is the axial dipole mass. In the charged current sector, where PCAC is assumed to be valid, the pseudoscalar form factor $g_{3}(Q^2)$, is dominated by the pion pole dominance of the divergence of the axial vector current~(PDDAC) and is given in terms of $g_{1}(Q^2)$ using the Goldberger-Trieman relation. However, $g_3(Q^2)$ does not contribute in the limit $m_l \rightarrow 0$. 
 As discussed above, the vector form factors are related to the electromagnetic form factors, which are determined precisely using the data from the real and virtual photon experiments on the nucleon targets, while the pseudoscalar form fIactor is related to the axial vector form factor. 
 Thus, one of main source of uncertainty in the quasielastic scattering cross sections is associated with the uncertainty in the axial vector form factor. 
 The value of axial vector charge $g_1(0)$ is determined precisely from the neutron decay experiments, and the value of $M_{A}$ is the major source of uncertainty as the various neutrino experiments report value of $M_{A}$ in the range 1--1.3~GeV~(see Table~2.2 of Ref.~\cite{SajjadAthar:2022pjt} for details).
 Therefore, to show the dependence of $\frac{d\sigma}{dQ^2}$ on $M_{A}$, in Fig.~\ref{fig:QE_MA}, we have presented the results for the $Q^2$-distribution at $E_{{\nu_\mu }({\bar\nu_\mu})}=2$~GeV i.e. $\frac{d\sigma}{dQ^2}$ vs $Q^2$, corresponding to the different values of $M_A$ i.e., $M_{A}= 1.026$, 1.1, 1.2 and 1.3~GeV, for the charged current $\nu_\mu$~(left panel) and $\bar\nu_\mu$~(right panel) induced quasielastic scattering processes on the neutron~($\nu_\mu + n \rightarrow \mu^- + p$) and proton~($\bar\nu_\mu + p \rightarrow \mu^+ + n$) targets, respectively. 
 It may be observed from the figure that the cross section increases with increase in the value of $M_{A}$, and this increase is more pronounced in the case of neutrino induced reaction than antineutrino induced reaction. 
 An increment in the value of $M_{A}$ by 30\% leads to an enhancement of about 25--35\% in the cross section in the region of $Q^2 \sim 0.4 - 0.6$~GeV$^2$, for both neutrino and antineutrino induced reactions.
 
In Figs.~\ref{fig:QE_g2R} and \ref{fig:QE_g2I}, we have presented the results for the $Q^2$-distribution at $E_{{\nu_\mu }({\bar\nu_\mu})}=2$~GeV for the charged current $\nu_\mu$~(left panel) and $\bar\nu_\mu$~(right panel) induced quasielastic scattering processes on the neutron~($\nu_\mu + n \rightarrow \mu^- + p$) and proton~($\bar\nu_\mu + p \rightarrow \mu^+ + n$) targets, respectively. The form of the induced tensor form factor $g_{2}(Q^2)$, which is associated with the second class currents, is taken to be of dipole form with $g_{2}(0)=\pm 3$ and same $M_A$ as for the $g_1(Q^2)$. The results presented in Fig.~\ref{fig:QE_g2R} are obtained to study the effect of second class currents assuming T-invariance by varying $g_{2}^{R}(0)$ between $0$ and $\pm 3$, while the results have been presented in Fig.~\ref{fig:QE_g2I} to study the effect of second class currents assuming time reversal violation by varying $g_{2}^{I}(0)$ between $0$ and $\pm 3$.

 It may be observed from the figure that in the presence of second class current assuming time reversal invariance~(Fig.~\ref{fig:QE_g2R}), which requires $g_{2} (Q^2)$ to be purely real i.e. $g_{2} (Q^2)= g_{2}^R (Q^2)$,  the cross 
 sections increases with the increase in the value of $g_{2}^{R} (0)$ and a change in the value of $g_{2}^{R} (0)$ from 0 to 3 results in an increase in the cross section by about $25\%$ at $Q^2 \sim 0.4 - 0.6$~GeV$^2$, for both neutrino and antineutrino induced reactions. We find that a change of sign in $g_{2}(0)$ i.e. $g_{2}(0)= -3$ gives almost the same result of the cross section as the one at $g_{2}(0)= +3$. 
 In the case of second class currents assuming time reversal violation, which requires $g_{2} (Q^2)$ to be purely imaginary i.e. $g_{2} (Q^2)= g_{2}^I (Q^2)$, a similar effect on the cross section~(Fig.~\ref{fig:QE_g2I}) is observed, as seen in the case of second class currents assuming time reversal invariance.

In Figs.~\ref{fig:QE_Y_g2R} and ~\ref{fig:QE_Y_g2I}, we have presented the results for the $Q^2$-distribution at $E_{\bar\nu_\mu}=2$~GeV for the charged current $\bar\nu_\mu$ induced $|\Delta S|=1$ quasielastic scattering processes on the proton~($\bar\nu_\mu + p \rightarrow \mu^+ + \Lambda$)~(left panel) and neutron~($\bar\nu_\mu + n \rightarrow \mu^+ + \Sigma^-$)~(right panel) targets, respectively. Assuming SU(3) symmetry, the $N-Y(=\Lambda, \Sigma)$ transition form factors have been taken to be of the similar form as taken in the case of the $|\Delta S|=0$ quasielastic scattering processes and discussed above. The results presented in Fig.~\ref{fig:QE_Y_g2R} are obtained in the presence of the second class currents assuming time reversal invariance by varying $g_{2}^{R}(0)$ between $0$ and $\pm 3$, while the results of T-violation studies have been presented in Fig.~\ref{fig:QE_Y_g2I} by varying $g_{2}^{I}(0)$ between $0$ and $\pm 3$. It may be observed from the figures that the non-zero value  of the form factor associated with the second class currents assuming time reversal invariance as well as its violation, increases the cross section, which is more pronounced in the case of $\Lambda$ production as compared to the results obtained for $\Sigma^-$ production. Unlike the case of $\Delta S=0$ quasielastic processes, in the case of $\Lambda$ production, we find that the cross section obtained using the negative value of $g_{2}^{R} (0)$ is larger than the one obtained using the positive value of $g_{2}^{R} (0)$. 
Quantitatively, in the case of $\Lambda$ production, we find an enhancement of about $25\%$~($40\%$) at $Q^2=0.5$~GeV$^{2}$ when $g_{2}^{R} (0)=+3~(-3)$ is taken in the numerical calculations, and this enhancement increases with the increase in the value of $Q^2$. It may be pointed out that the $\Lambda$ production is more suitable for the SCC or T-violation studies as their effect on $\Lambda$ polarization can be studied by analyzing the angular distribution of pions produced in $\Lambda \rightarrow p \pi^-$ decays~\cite{SajjadAthar:2022pjt}. It has been observed~(not shown here)~\cite{Fatima:2018tzs} that the presence of SCC with or with T invariance increases the total cross section. Moreover, a higher value of $M_{A}$ also leads to an increase in the cross section. 
Presently, in the neutrino physics community, there is an ongoing debate since the last decade to fix the value of $M_{A}$ as the results of the different experiments require different values of $M_{A}$ in the Monte Carlo to explain the data. 
In this light of this discussion, we have proposed that the presence of second class currents, if any, would help in fixing the value of $M_{A}$~\cite{Fatima:2018tzs}. 
Furthermore, we have also studied the polarization of the produced hyperons~\cite{Fatima:2018tzs} and have shown that the transverse component of the hyperon polarization, which lies perpendicular to the reaction plane, would be nonzero only in the presence of the SCC with T violation. 
Since, the DUNE experiment would use LArTPC type of detector, where the three-dimensional image of the interaction vertex can be reconstructed, therefore, at DUNE the study of transverse component of the polarized hyperon is feasible.
To conclude, if it becomes possible to put some limit on the transverse component of $\Lambda$ polarization in the future, then the information on the SCC can be obtained and using this information on the SCC form factor, it becomes possible to contraint the value of $M_{A}$ for the neutrino experiments.

The results for the $\Lambda$ production cross sections from the free nucleons
as a function of antineutrino energies are presented in Fig.~\ref{free_hyperon}, at the two values of $M_A$ viz. $M_A = 1.026$ GeV and 1.2 GeV. It may be observed from these figures that the cross section increases with energy and the increase is about 5$\%$ at $E_{\bar{\nu}_\mu} = 
1$ GeV. A comparison is made with available experimental results from 
CERN~\cite{Eichten:1972bb, Erriquez:1977tr, Erriquez:1978pg}, BNL~\cite{Fanourakis:1980si}, FNAL~\cite{Ammosov:1986jn, 
Ammosov:1986xv} and SKAT~\cite{SKAT:1989nel} experiments as well as with the theoretical calculations performed by Wu et 
al.~\cite{Wu:2013kla} and Finjord and Ravndal~\cite{Finjord:1975zy} using quark model and the calculations performed by 
Erriquez et al.~\cite{Erriquez:1978pg}, Brunner et al.~\cite{SKAT:1989nel} and Kuzmin and Naumov~\cite{Kuzmin:2008zz} based 
on the prediction using Cabibbo theory. 
Although the theoretical results obtained using the different models show significant variation among themselves, a reasonable agreement of our results using $M_{A}=1.026$~GeV with the experimental results can be seen. However, a precise determination of the experimental cross sections of the quasielastic hyperon production induced by the antineutrinos is needed.

 \subsection{Inelastic scattering processes}
 \begin{figure}  
\centering
\includegraphics[width=16cm,height=8cm]{cc_comparison.eps}
\caption{$\sigma$ vs. $E_{\nu_{\mu}}$ for the processes $\nu_{\mu} p \longrightarrow \mu^{-}   p   \pi^{+}$~(left panel) and
$\nu_{\mu}   n \longrightarrow \mu^{-}   p   \pi^{0}$~(right panel) with $W<1.4$~GeV. Solid line is the result of the present model with deuteron effect; compared with 
other theoretical models like DCC~\cite{Nakamura:2015rta}~(dashed line), HNV~\cite{Hernandez:2007qq}~(dashed-dotted line), 
Hybrid~\cite{Gonzalez-Jimenez:2016qqq}~(double-dotted-dashed line), 
NuWro~\cite{Gonzalez-Jimenez:2016qqq}~(double-dashed-dotted line) and Giessen~\cite{Lalakulich:2010ss}~(dotted line). Data points quoted as ANL extracted and BNL extracted are the 
reanalyzed data by Wilkinson et al.~\cite{Wilkinson:2014yfa} and Rodrigues et al.~\cite{Rodrigues:2016xjj}. }\label{fig:cc_compare}
\end{figure}

  \begin{figure}
 \begin{center}
    \includegraphics[height=8cm,width=7.5cm]{dsig_dq2_Kaon.eps}
    \hspace{3mm}
      \includegraphics[height=8cm,width=7.5cm]{dsig_dq2_antiKaon.eps} 
       \end{center}
  \caption{$\frac{d\sigma}{dQ^2}$ vs. $Q^2$ for the single kaon production processes.}
   \label{fig:Kaon}
 \end{figure} 
 
   \begin{figure}
 \begin{center}
    \includegraphics[height=8cm,width=7.5cm]{dsigma_dq2_eta_neutrino.eps}
    \hspace{3mm}
      \includegraphics[height=8cm,width=7.5cm]{dsigma_dq2_eta_antineutrino.eps} 
       \end{center}
  \caption{$\frac{d\sigma}{dQ^2}$ vs. $Q^2$ for the processes $\nu_{\mu} + n \longrightarrow \mu^{-} + p + \eta$~(left panel) and $\bar{\nu}_{\mu} + p \longrightarrow \mu^{+} + n + \eta$~(right panel) at $E_{\nu}=2$~GeV. 
  Solid~(dashed-dotted) line represents the results obtained using the full model when a cut on $W<2$~GeV~($W<1.6$~GeV) is applied. 
  Dashed~(dotted) line represents the results obtained using the background and $S_{11}(1535)$ resonance, when a cut on $W<2$~GeV~($W<1.6$~GeV) is applied.}
   \label{fig:eta}
 \end{figure}
 
   \begin{figure}
 \begin{center}
    \includegraphics[height=8cm,width=7.5cm]{dsigma_dq2_associated_production_nu.eps}
    \hspace{3mm}
      \includegraphics[height=8cm,width=7.5cm]{dsigma_dq2_associated_production_antinu.eps} 
       \end{center}
  \caption{$\frac{d\sigma}{dQ^2}$ vs. $Q^2$ for the processes $\nu_{\mu} + n \longrightarrow \mu^{-} + \Lambda + K^{+}$~(left panel) and $\bar{\nu}_{\mu} + p \longrightarrow \mu^{+} + \Lambda + K^0$~(right panel) at $E_{\nu}=2$~GeV. 
  Solid~(dashed-dotted) line represents the results obtained using the full model when a cut on $W<2$~GeV~($W<1.8$~GeV) is applied. 
  }
%   Dashed~(dotted) line represents the results obtained using the background and $S_{11}(1535)$ resonance, when a cut on $W<2$~GeV~($W<1.6$~GeV) is applied.}
   \label{fig:associated}
 \end{figure}

The results are presented in Fig.~\ref{fig:pi} for $\frac{d\sigma}{dQ^2}$ vs $Q^2$ for the charged current neutrino and antineutrino induced one pion production off the nucleon target. These results are presented on the left panel for $\nu_{\mu}   p 
\longrightarrow \mu^{-}   p   \pi^{+}$~(top panel), $\nu_{\mu}   n \longrightarrow \mu^{-}   n  \pi^{+}$~(central panel), 
$\nu_{\mu}   n \longrightarrow \mu^{-}   n   \pi^{0}$~(bottom panel), and on the right panel for $\bar\nu_{\mu}   p 
\longrightarrow \mu^{+}   n   \pi^{-}$~(top panel), $\bar\nu_{\mu}   n \longrightarrow \mu^{+}   p  \pi^{-}$~(central panel), 
$\bar\nu_{\mu}   p \longrightarrow \mu^{+}   n   \pi^{0}$ processes. In the case of single pion production, we have taken the contribution from 
spin $\frac{1}{2}$ resonances like $P_{11} (1440)$, $S_{11} (1535)$, $S_{31} (1620)$, $S_{11} (1650)$, and spin $\frac{3}{2}$ resonances like $P_{33} (1232)$, $D_{13} (1520)$, $D_{33} (1700)$, $P_{13} (1720)$. In the vector sector, the helicity amplitudes for all these resonance excitations are given by the MAID parameterization~ \cite{Tiator:2011pw}. In the case of spin $\frac{1}{2}$ resonances, the s-channel and u-channel hadronic  currents for the positive and negative parity resonances are discussed in detail in Ref.~\cite{SajjadAthar:2022pjt}. We have also considered the nonresonant background contribution~\cite{SajjadAthar:2022pjt}. The Full model corresponds to the results when Eq.~(\ref{allterms}) is used in the evaluation of the matrix element, while $P_{33}(1232)$ corresponds to the results when only $P_{33}(1232)$ dominance is assumed. The results are presented with a cut of 1.4~GeV and 2.0~GeV on the center of mass energy $W$, while calculating the differential scattering cross section. 
It may be observed from the figure that in the case of $p\pi^{+}$ and $n\pi^{-}$ channels, which are purely isospin $I=\frac{3}{2}$ states, the cross section is dominated mainly by $P_{33}(1232)$ resonance and the contribution from the nonresonant terms as well as from the higher resonances is almost negligible in the kinematic region considered in this work. 
The overall cross section decreases when a cut of $W<1.4$~GeV is applied and in the peak region of $Q^2$, the cross section is decreased by about $23\%~(27\%)$ as compared to the results obtained when a cut of 2~GeV on $W$ is applied, for $p\pi^{+}~(n\pi^{-})$ channels. 
In the case when a cut of $W<1.4$~GeV is applied, the results obtained using the full model vs. considering only $P_{33}(1232)$ contribution are almost the same as there is no resonance other than $P_{33}(1232)$ in the range of $W\sim 1.4$~GeV contributing to the pion production channel. 
In the case of $n\pi^{+}$ and $p\pi^{-}$ channels, we find some contribution from the nonresonant terms as well as from the higher resonances, when a cut of $W<2$~GeV is applied. 
A cut of $W<1.4$~GeV reduces the cross section by about $75\%~(78\%)$ for the $n\pi^{+}~(p\pi^{-})$ channels as compared to the results obtained when a cut of 2~GeV is applied on the center of mass energy.
It may be observed from the figure that the presence of the nonresonant terms and the higher resonances leads to an enhancement in the cross section, which is about 15--20\% in the peak region of $Q^2$, 
in the case of $p\pi^0$ and $n\pi^0$ channels. 
Moreover, we observe a suppression in the cross section, which is about $50\%~(55\%)$ for the $p\pi^{0}~(n\pi^{0})$ channels when a cut of 1.4~GeV is applied instead of a cut of 2~GeV on the center of mass energy $W$.

In Fig.~\ref{fig:cc_compare}, we have compared the theoretical results for the single pion production induced by neutrinos 
when a cut of 1.4~GeV is applied on $W$, obtained in the different models like the present 
model, the dynamical coupled channel~(DCC) model by Nakamura et al.~\cite{Nakamura:2015rta}, 
the Hernandez, Nieves, and Valverde~(HNV) model~\cite{Hernandez:2007qq}, the extension of HNV model by incorporating Regge model at high 
energies~(Hybrid) by Gonzalez-Jimenez et al.~\cite{Gonzalez-Jimenez:2016qqq}, the results from 
NuWro~\cite{Gonzalez-Jimenez:2016qqq} Monte Carlo generator, and the Giessen model by Lalakulich et 
al.~\cite{Lalakulich:2010ss}. It may be observed from the figure that in the case of $p\pi^+$ channel, the results obtained in 
our model are quite consistent with the results obtained by the hybrid model and are in a very good agreement with the 
reanalyzed data of ANL~\cite{Radecky:1981fn} and BNL~\cite{Kitagaki:1986ct} by Wilkinson et al.~\cite{Wilkinson:2014yfa} and 
Rodrigues et al.~\cite{Rodrigues:2016xjj}. However, the results obtained in the other models like DCC, HNV, etc., are higher 
than the results obtained by us as well as the experimental data, but are consistent with one another. Moreover, the 
results obtained by the Giessen group~\cite{Lalakulich:2010ss} are lower than our results. In the case of $p\pi^0$ channel, 
our results are in a quite good agreement with the experimental data, while the results obtained in the other  theoretical 
models are higher than our results. At energies $E_{\nu_{\mu}}<0.8$~GeV, the results obtained in the various models are 
consistent with each other. 
It may be noticed 
from the figure that there is a large difference among the various theoretical models and Monte Carlo generators available in 
the literature. In order to understand the dynamics of the single pion production, which is the simplest inelastic process, 
further theoretical and experimental work is required.

In Fig.~\ref{fig:Kaon}, the results are presented for $\frac{d\sigma}{dQ^2}$ vs $Q^2$ for the charged current $\nu_\mu$~(left panel) and $\bar\nu_\mu$~(right panel) induced one kaon production off the nucleon target. These results are presented on the left panel for the processes $\nu_{\mu}   N 
\longrightarrow \mu^{-}   N^\prime   K^{i};~~N, N^\prime=n$ or $p$; $K^{i}= K^+$ or $K^0$, where only the nonresonant background contributions have been considered as there is no established resonance which decays to a kaon and a nucleon in the region of $W < 2$~GeV. 
On the right panel, the results are presented for the processes $\bar\nu_{\mu}   N 
\longrightarrow \mu^{+}   N^\prime   K^{i};~~N, N^\prime=n$ or $p$; $K^{i}= K^-$ or $\bar K^0$, we have taken the contribution from the nonresonant terms as well as from the $\Sigma^{\star}(1385)$ ($S=-1$) resonance. The results are presented with a cut of 1.6~GeV and 2.0~GeV on $W$, while calculating the differential scattering cross section. It may be observed from the figure that the effect of $W$ cut is quite important for the single kaon production processes. 
In the case of neutrino induced single kaon production, when the cut on $W$ is decreased from 2~GeV to 1.6~GeV, the cross section for all three processes decreases by about 75--80\%, while in the case of antineutrino induced antikaon production processes, this reduction is about 60--65\% for the three different reactions viz. $\bar{\nu}_{\mu} + p \longrightarrow \mu^{+} + p + K^{-}$, $\bar{\nu}_{\mu} + p \longrightarrow \mu^{+} + n + \bar{K}^{0}$ and $\bar{\nu}_{\mu} + n \longrightarrow \mu^{+} + n + K^{-}$.
 
         \begin{figure}
 \begin{center}
    \includegraphics[height=8cm,width=7.5cm]{f2_free_weak_pdfs.eps}
%         \hspace{3mm}
            \includegraphics[height=8cm,width=7.5cm]{f3_free_pdfs.eps}
       \end{center}
  \caption{Neutrino-nucleon structure functions $F_{2N}(x,Q^2)$ (left panel) and $x F_{3N}(x,Q^2)$ (right panel) vs. $x$ at $Q^2=2$~GeV$^2$ and 10 GeV$^2$. 
   These results are obtained at NLO using different PDFs parameterizations~\cite{Gluck:2007ck, Nadolsky:2008zw, Martin:2009iq, Harland-Lang:2014zoa, Dulat:2015mca}.}
   \label{fig:dissf}
 \end{figure}
 
Fig.~\ref{fig:eta} presents the results for $\frac{d\sigma}{dQ^2}$ vs $Q^2$ for the charged current $\nu_\mu$~(left panel) and $\bar\nu_\mu$~(right panel) induced single eta production off the nucleon target. These results are presented on the left panel for $\nu_{\mu}   n \longrightarrow \mu^{-}   p   \eta$ and on the right panel for $\bar\nu_{\mu}   p \longrightarrow \mu^{+}   n   \eta$. We have taken the contribution  from the spin $\frac{1}{2}$ resonances like $S_{11} (1535)$, $S_{11} (1650)$, $P_{11}(1710)$, $P_{11}(1880)$, and $S_{11}(1895)$ along with the s-channel and u-channel Born terms. The $\eta$ production is dominated by the excitation of  $S_{11}(1535)$ resonance and its subsequent decay to $\eta N$. 
To show the dominance of $S_{11}(1535)$ resonance in the case of $\eta$ production, we have also presented the results by taking into account the contribution from $S_{11}(1535)$ resonance only.
The results are presented with a cut of 1.6~GeV and 2.0~GeV on $W$, while calculating the differential scattering cross section. From the figure, one may observe  that in the case of $W<1.6$~GeV cut, the results obtained with the full model and considering only $S_{11}(1535)$ resonance, overlap, as there is no other resonance contributing in this region of $W$ and the contribution of the nonresonant terms in the case of $\eta$ production is very small. 
As the cut on $W$ is increased from 1.6~GeV to 2~GeV, the differential cross section increases, and in the peak region of $Q^2$, this increase in the cross section is about 35--40\% for both the neutrino and antineutrino induced $\eta$ production processes.
Recently, MicroBooNE collaboration~\cite{MicroBooNE:2023ubu} has reported the flux-averaged cross section for the neutrino induced $\eta$ production from the Argon target to be $<\sigma> = (3.22 \pm 0.84 \pm 0.86) \times 10^{-41}$~cm$^{2}/$nucleon. 
When the interaction of a neutrino takes place from the nucleons bound inside a nucleus, the nuclear medium and final state interaction effects become important. 
To compare our results with the MicroBooNE data, we have extended our model to study the production of $\eta$ meson from the Argon nuclear target by taking into account the nuclear medium effect and width modification of $S_{11}(1535)$ resonance, and found the flux-averaged cross section to be $<\sigma> = 1.78\times 10^{-41}$~cm$^{2}/$nucleon. However, the study of final state interaction of the produced $\eta$ from the residual nucleus is ongoing and we will report those results elsewhere, and it is expected that the effect of final state interaction would further reduce the cross section. 
Comparing our results with the MicroBooNE data, we find that the experimental data show significant difference from our theoretical results. Thus, more experimental and theoretical studies are needed. 
 
The results are presented in Fig.~\ref{fig:associated} for $\frac{d\sigma}{dQ^2}$ vs $Q^2$ for the charged current $\nu_\mu$~(left panel) and $\bar\nu_\mu$~(right panel) induced associated particle production off the nucleon target. These results are presented on the left panel for $\nu_{\mu}   n 
\longrightarrow \mu^{-}   \Lambda   K^{+}$ process and on the right panel for $\bar\nu_{\mu}   p \longrightarrow \mu^{+}   \Lambda   K^{0}$ process. We have taken the contribution from the nonresonant terms as well as $S_{11} (1650)$, $P_{11}(1710)$, and $P_{13}(1720)$ resonances. The results are presented with a cut of 1.8~GeV and 2.0~GeV on $W$, while calculating the differential scattering cross section. 
It may be observed that there is a large suppression due to the $W<2$~GeV cut in the peak region of the $Q^2$ distribution.

 \subsection{Deep inelastic scattering}
   \begin{figure}
 \begin{center}
    \includegraphics[height=8.5cm,width=7.5cm]{dsigdq2_3gev_numu.eps}
    \hspace{3mm}
      \includegraphics[height=8.5cm,width=7.5cm]{dsigdq2_3gev_numubar.eps} 
       \end{center}
  \caption{$\frac{d\sigma}{dQ^2}$ vs. $Q^2$ for the deep inelastic scattering processes induced the neutrino~(left panel) and antineutrino~(right panel) at $E_{\nu}=3$~GeV. 
   Solid~(dashed) line represents the results obtained using NLO, TMC, and HT effects,  when a cut on $W \ge 2$~GeV~($W \ge 1.8$~GeV) is applied. 
  Double-dotted-dashed~(dashed-dotted) line represents the results obtained using NLO and TMC, when a cut on $W \ge 2 $~GeV~($W \ge 1.8$~GeV) is applied.
  }
   \label{fig:dis}
 \end{figure}
 
     \begin{figure}
 \begin{center}
    \includegraphics[height=8cm,width=10.5cm]{dsigy_cc_numu.eps}
       \end{center}
  \caption{$\frac{d\sigma}{dy}$ vs. $y$ for the deep inelastic scattering processes induced the neutrino~(solid lines) and antineutrino~(dashed lines)
  at $E_{\nu}=160$~GeV. 
   These results are obtained using NLO, TMC, and HT effects, and are compared with the experimental data from CHARM collaboration~\cite{CHARM:1981zpm}.
  }
   \label{fig:disx}
 \end{figure}
 
 In Fig.~\ref{fig:dissf}, we present the results for the free nucleon structure functions~(Eq. 37), $F_{2N}(x,Q^2)$ (left panel) and $xF_{3N}(x,Q^2)$ (right panel)
 as functions of  $x$, obtained using various 
PDF parameterizations, like GJR~\cite{Gluck:2007ck}, CTEQ6.6~\cite{Nadolsky:2008zw}, MSTW~\cite{Martin:2009iq}, MMHT~\cite{Harland-Lang:2014zoa} 
and CT14~\cite{Dulat:2015mca} at $Q^2=2$~GeV$^{2}$ and 10~GeV$^2$. 
This comparison highlights the dependence of nucleon structure functions on the choice of PDF parameterizations.  The structure functions were evaluated at the next-to-leading order~(NLO) following the methodologies prescribed by Vermaseren et al.~\cite{Vermaseren:2005qc} and Moch et al.~\cite{Moch:2008fj}. From the left panel, it is evident that the results obtained using MSTW, MMHT, and CT14 parameterizations are in close agreement across the entire range of $x$ for both $Q^2$ values. However, notable deviations are observed in the low to intermediate $x$ region~($x \le$ 0.3) when comparing these results with those derived from the CTEQ6.6~\cite{Nadolsky:2008zw} and GJR~\cite{Gluck:2007ck} parameterizations. These numerical results become consistent among themselves with the increase in $x$ and $Q^2$.
From the right panel of the figure, it may be noticed that the results 
of $xF_{3N}(x,Q^2)$ obtained by using GJR~\cite{Gluck:2007ck} and MMHT~\cite{Harland-Lang:2014zoa} parameterizations 
are significantly different from the results obtained by
using CTEQ6.6~\cite{Nadolsky:2008zw}, MSTW~\cite{Martin:2009iq} and CT14~\cite{Dulat:2015mca} parameterizations in the region of low to intermediate 
$x$ and low $Q^2$. However, at $Q^2=10$ GeV$^2$ this difference becomes small and the numerical results for different PDFs parameterizations are
found to be in good agreement in the entire range of $x$, except for the results using GJR~\cite{Gluck:2007ck} parameterization in the region of
low to intermediate $x$. This uncertainty arising due to the PDFs dependence of nucleon structure functions would further translate in the determination of 
neutrino-nucleon scattering cross sections in the kinematic region of low to intermediate $x$ and at the moderate values of $Q^2$. Hence, parton 
distribution functions are needed to be well determined in this kinematic region of $x$ and $Q^2$.

In Fig.~\ref{fig:dis}, we have presented the results for $\frac{d\sigma}{dQ^2}$ vs $Q^2$ for the 
charged current $\nu_\mu$ (left panel) and $\bar\nu_\mu$ (right panel) induced deep inelastic 
scattering off the nucleon target at $E_{{\nu_\mu }{\bar\nu_\mu}}$=3 GeV. This is because 
considering $W >1.8$ or $2$ GeV at $E_{{\nu_\mu }{(\bar\nu_\mu)}}$=2 GeV would result a 
small contribution in the lower region of $Q^2$ only. These numerical results are obtained 
by incorporating the higher order perturbative corrections up to next-to-the-leading order~(NLO) 
as well as the nonperturbative effects like target mass correction~(TMC) and higher twist~(HT) correction up to twist-4. These higher order perturbative and nonperturbative
corrections are discussed in detail in Ref.~\cite{Zaidi:2019asc, Ansari:2020xne}. It may be noticed that the numerical results obtained with a lower cut on $W$, i.e., $W>1.8$ GeV span over the wider kinematic range of $Q^2$ as well as are higher than the results obtained with a cut of $W>2$ GeV. For example, the neutrino-nucleon (antineutrino-nucleon) differential cross section with a cut of $W>2$ GeV gets suppressed by 40\% (44\%) at $Q^2=1$ GeV$^2$, 45\% (50\%) at $Q^2=1.5$ GeV$^2$ and 60\% (64\%) at $Q^2=2$ GeV$^2$ as compared to the results obtained with $W>1.8$ GeV cut. It may be observed that the effect of center of mass energy cut is more pronounced in the peak region as well as it excludes the region of high $x$ relevant to the tail of the differential cross sections, where the nonperturbative effects are significant. Furthermore, we find that the differential cross section gets reduced due to the inclusion of higher twist corrections in the entire range of $Q^2$. This reduction in the differential cross section is more pronounced for the lower values of $Q^2$ which gradually becomes small with the increase in $Q^2$, for example, there is a reduction of about 16\% at $Q^2=1$ GeV$^2$, 14\% at $Q^2=1.5$ GeV$^2$ and 12\% at $Q^2=2$ GeV$^2$ in the $\nu_\mu$ induced deep inelastic cross sections. It is important to point out that the inclusion of higher twist~(HT) correction in $\bar \nu_\mu-N$ deep inelastic cross sections
results in an enhancement, which is contrary to the behavior of HT correction observed in the case of $\nu_\mu-N$ DIS reaction. Quantitatively, this enhancement in the numerical results incorporating a cut of $W>2$ GeV is about 14\% at $Q^2=1$ GeV$^2$ which increases to 30\% at $Q^2=1.5$ GeV$^2$.

  In Fig.~\ref{fig:disx}, the numerical results are presented for $\frac{1}{E_\nu}\frac{d\sigma}{dy}$ vs $y$ at $E_\nu=160$ GeV for the charged current 
  $\nu_\mu-N$ (solid line) and $\bar\nu_\mu-N$ (dashed line) deep inelastic scattering cross sections. These numerical results are obtained at NLO incorporating
  the nonperturbative corrections like TMC following the operator product expansion approach and higher twist correction up to twist-4 following the 
  renormalons approach~\cite{Zaidi:2019asc, Ansari:2020xne}. We have found that at this (anti)neutrino energy these nonperturbative corrections 
  are almost negligible (not shown here explicitly). 
  The antineutrino-nucleon cross sections are lower compared to the neutrino-nucleon cross sections, with the degree of suppression increasing as $y$ increases. As observed from the figure, the numerical results align well with the CHARM experimental data~\cite{CHARM:1981zpm}, except in the peak region where slight deviations are evident.

\section{Summary and conclusions}
\label{summary}
We have studied the charged current induced neutrino and antineutrino scattering from the free nucleon target in the few GeV energy region, relevant
for the accelerator and atmospheric neutrino experiments, where the contribution to the cross section mainly comes from the
quasielastic, inelastic one pion, one kaon, one eta, single hyperon, associated particle production, and the deep inelastic
scattering processes. In the case of quasielastic scattering and the single hyperon production induced by antineutrinos off the
nucleon target, we have also studied the effect of second class current. For the one pion production, one eta production, single 
kaon production and associated particle production, we have taken the contribution of nonresonant as well as resonant contributions up to $W < 2$ GeV.
The numerical results are presented for the $Q^2$ distribution of the differential cross section for all the aforementioned processes.
The effect of the cut on the center of mass energy $W$ has been explicitly discussed. 

We find that:
\begin{itemize}

\item The quasielastic scattering cross sections dominate at low (anti)neutrino energies and make significant contributions even at $E_\nu=2$ GeV. The cross sections are sensitive to the axial dipole mass $M_A$ and are used to phenomenologically determine the value of $M_A$.

 \item The cross sections are found to be sensitive to the presence of second class currents(SCC) with (without) T-invariance if the strength of the real (imaginary) value to the weak electric form factor $g_2^R$($g_2^I$) is large in the range of $g_2^R$ ($g_2^I$)$\simeq 2-3$. The change in the sign of  the 
 strength of $g_2^R$ has little effect on the numerical results for the cross sections.
 
 \item It is found that the presence of SCC with or without T-invariance increases the cross sections, an effect which is achieved by a larger $M_A$. 
 Therefore, the presence or absence of SCC plays an important role in the  phenomenological determination of $M_A$, which is generally done by assuming no second 
 class currents.
 
 \item The results for the antineutrino cross sections are qualitatively similar but smaller as compared to the neutrino scattering cross sections. 
 However, in this case the additional channel of hyperon production $\Lambda$, $\Sigma$ is opened which is very useful in doing SCC and T-invariance studies
 as the effects are larger and can be pursued by studying the angular distribution of pions coming from $\Lambda(\Sigma) \to N\pi$ decays. 
 Moreover, the study of SCC is quite relevant in a broader manner as the presence of SCC would directly imply beyond the standard model physics. 
 The collider experiments like CMS~\cite{CMS:2018wjk}, Belle~\cite{Belle:2018ttu}, LHCb~\cite{LHCb:2013hzx}, etc. have reported the results for the transverse component of $\Lambda$, $\Lambda_b$, $\Lambda_{c}$ hyperons, which is the direct evidence of physics beyond the standard model. 
 Therefore, our results on the weak production of $\Lambda$ hyperons and its polarization observables would provide an alternative method to study physics beyond the standard model.
 
 \item CCQE cross section is followed by pion production cross section, which is dominated by $p\pi^+$ final state in the case of neutrino induced reaction due to $\Delta^{++}$ dominance. The nonresonant background terms include five diagrams viz. direct and cross nucleon pole, contact term, pion pole and pion in flight terms~\cite{SajjadAthar:2022pjt}. For the $\Delta(1232)$ resonance we have included both direct and cross diagrams. Apart from the $\Delta(1232)$ resonance, which mainly decays to $N\pi$, we have also taken contributions from $P_{11}$(1440), $S_{11}$(1535), 
$S_{31}(1620)$, and $S_{11}(1650)$ spin half resonances and $D_{13}(1520)$, $D_{33}(1700)$, and $P_{13}(1720)$ spin three-half resonances and considered both s-channel and u-channel contributions~\cite{SajjadAthar:2022pjt}.  $p\pi^0$ and $n\pi^+$ final states  have significant contribution from the nonresonant as well as other resonance states. In the case of antineutrino induced $n\pi^-$, $p\pi^-$ and $n\pi^0$ production, $n\pi^-$ dominates due to $\Delta^{-}$ resonant state. $p\pi^-$ and $n\pi^0$ production cross sections are almost equal orders of magnitude. 

 \item In case of the inelastic processes in which other mesons like $\eta$ and $K$ are produced by the $\Delta S=0$ and $\Delta S=1$ currents, the 
 cross sections are about more than one order of magnitude smaller than the quasielastic and one pion production at $E_{\nu(\bar\nu)}\approx 2$ GeV.
 In case of $\Delta S=0$ production, the cross section for the $\eta$ and $\Lambda K$ particles are suppressed by the threshold effect while in 
 the case of $K$ production, through the $\Delta S=1$ currents, the cross sections are reduced by the Cabibbo suppression.
 
 \item $\eta$ production is dominated by the $S_{11}(1535)$ resonance while the $\Delta S=1$ $K$ production is dominated by the 
 contact term in the nonresonance contribution. In the case of associated production of $\Lambda K$ through the $\Delta S=0$ currents, no single resonance dominates 
 but the contribution of $P_{11}(1710)$ in the neutrino reactions and the contribution of $P_{13}(1720)$ in the case of antineutrino reactions are significant.
 
 \item The deep inelastic cross sections with a cut of $E_{\nu(\bar\nu)}= 2$ GeV makes significant contributions at $E_{\nu(\bar\nu)}\approx 2.5$ GeV and becomes 
 larger than the inelastic cross sections for $E_{\nu(\bar\nu)}\approx 3$ GeV.
 
 \item In the case of DIS, it is found that the higher order perturbative and nonperturbative corrections has significant effect on the cross sections.
 
 \item The effect of CM energy cut of $W>1.8$ GeV or $W>2$ GeV makes a significant difference on the cross sections. The MINER$\nu$A collaboration is currently analysing 1D and 2D results on the DIS of (anti)neutrino-nucleus scattering for the different nuclear targets like carbon, hydrocarbon, iron and lead. Therefore, it is very important to first understand what cut
 on $W$ would be appropriate to identify the true DIS events in the (anti)neutrino-nucleon scattering. 
\end{itemize}

The results discussed in this paper will be quite useful to estimate the $\nu_l(\bar\nu_l)-N$ scattering cross sections, particularly for the experiments, which are studying $Q^2$-distribution in the few GeV energy region.
The MicroBooNE collaboration has reported the cross section results for the single hyperon production and $\eta$ meson production processes and plans to study the single kaon production and associated particle production processes in the near future. 
Moreover, the DUNE experiment also plans to perform cross section measurements in the neutrino energy region of about 2~GeV. 
Therefore, the results discussed in this paper would be useful in the analysis of the accelerator neutrino experiments like MicroBooNE, NOvA, DUNE, etc., as well as the other experiments with atmospheric neutrinos.

 \section*{Acknowledgments}
MSA and AF are thankful to the
Department of Science and Technology (DST), Government of India for providing financial assistance under Grant No.
SR/MF/PS-01/2016-AMU. FZ is thankful to Council of Scientific $\&$ Industrial Research, Govt. of India for providing Senior Research Associateship (SRA) under the Scientist’s Pool Scheme, file no. 13(9240 A)2023 POOL.

%Unused bibitems


\begin{thebibliography}{100}
\bibitem{DUNE:2018tke}
B.~Abi \textit{et al.} [DUNE],
[arXiv:1807.10334 [physics.ins-det]].
 


\bibitem{DUNE:2024wvj}
A.~Abed Abud \textit{et al.} [DUNE],
[arXiv:2408.12725 [physics.ins-det]].
 
 


\bibitem{Hyper-Kamiokande:2018ofw}
K.~Abe \textit{et al.} [Hyper-Kamiokande],
[arXiv:1805.04163 [physics.ins-det]].
 
%\cite{Cerrone:2024vvk}
\bibitem{Cerrone:2024vvk}
V.~Cerrone [JUNO],
%``Current status and physics prospects of the JUNO experiment,''
Nuovo Cim. C \textbf{47}, no.3, 70 (2024)
doi:10.1393/ncc/i2024-24070-7
%0 citations counted in INSPIRE as of 09 Dec 2024

\bibitem{KATRIN:2022ayy}
M.~Aker \textit{et al.} [KATRIN],
J. Phys. G \textbf{49}, no.10, 100501 (2022)
doi:10.1088/1361-6471/ac834e
[arXiv:2203.08059 [nucl-ex]].


\bibitem{LSND:1996ubh}
C.~Athanassopoulos \textit{et al.} [LSND],
Phys. Rev. Lett. \textbf{77}, 3082-3085 (1996).

\bibitem{LSND:2001aii}
A.~Aguilar \textit{et al.} [LSND],
Phys. Rev. D \textbf{64}, 112007 (2001).


\bibitem{MiniBooNE:2007uho}
A.~A.~Aguilar-Arevalo \textit{et al.} [MiniBooNE],
Phys. Rev. Lett. \textbf{98}, 231801 (2007).

\bibitem{MiniBooNE:2008yuf}
A.~A.~Aguilar-Arevalo \textit{et al.} [MiniBooNE],
Phys. Rev. Lett. \textbf{102}, 101802 (2009)
doi:10.1103/PhysRevLett.102.101802
[arXiv:0812.2243 [hep-ex]].






\bibitem{Denton:2021czb}
P.~B.~Denton,
Phys. Rev. Lett. \textbf{129}, 061801 (2022).


\bibitem{IceCube:2024dlz}
R.~Abbasi \textit{et al.} [IceCube],
Phys. Rev. D \textbf{110}, 072007 (2024).

\bibitem{STEREO:2022nzk}
H.~Almaz\'an \textit{et al.} [STEREO],
Nature \textbf{613}, 257 (2023).


\bibitem{IceCube:2016xxt}
M.~G.~Aartsen \textit{et al.} [IceCube],
J. Phys. G \textbf{44}, no.5, 054006 (2017)
doi:10.1088/1361-6471/44/5/054006
[arXiv:1607.02671 [hep-ex]].

\bibitem{Ferrara:2024try}
G.~Ferrara [KM3NeT],
Nuovo Cim. C \textbf{47},  71 (2024).


\bibitem{SajjadAthar:2022pjt}
M.~Sajjad Athar, A.~Fatima and S.~K.~Singh,
Prog. Part. Nucl. Phys. \textbf{129}, 104019 (2023).
  

\bibitem{Athar:2020kqn}
M.~Sajjad~Athar and S.~K.~Singh,
Cambridge University Press, 2020,
ISBN 978-1-108-77383-6, 978-1-108-48906-5
doi:10.1017/9781108489065

\bibitem{SajjadAthar:2021prg}
M.~Sajjad Athar, S.~W.~Barwick, T.~Brunner, J.~Cao, M.~Danilov, K.~Inoue, T.~Kajita, M.~Kowalski, M.~Lindner and K.~R.~Long, \textit{et al.}
Prog. Part. Nucl. Phys. \textbf{124}, 103947 (2022)
doi:10.1016/j.ppnp.2022.103947
[arXiv:2111.07586 [hep-ph]].

\bibitem{Ruso:2022qes}
L.~A.~Ruso, A.~M.~Ankowski, S.~Bacca, A.~B.~Balantekin, J.~Carlson, S.~Gardiner, R.~Gonz\'alez-Jim\'enez, R.~Gupta, T.~J.~Hobbs and M.~Hoferichter, \textit{et al.}
[arXiv:2203.09030 [hep-ph]].


\bibitem{Fermi:1934sk} 
  E.~Fermi,
    Nuovo Cim.\  {\bf 11}, 1 (1934).
  
  

\bibitem{Perrin:1933} 
F. Perrin, in Structure et Proprietes des Noyaux Atomiques, Rapports et Dis-
cussions du Septiem e Conseil de Physique, idem p. 327; Compt. Rendus 197,
1625 (1933).

\bibitem{Gamow:1936jk} 
  G.~Gamow and E.~Teller, Phys.\ Rev.\  {\bf 49}, 895 (1936).
  
  

\bibitem{Lee:1956vjd} 
  T.~D.~Lee and C.~N.~Yang, Phys.\ Rev.\  {\bf 102}, 290 (1956).
  

\bibitem{Lee:1956qn} 
  T.~D.~Lee and C.~N.~Yang, Phys.\ Rev.\  {\bf 104}, 254 (1956).


\bibitem{Wu:1957my} 
  C.~S.~Wu, E.~Ambler, R.~W.~Hayward, D.~D.~Hoppes and R.~P.~Hudson, Phys.\ Rev.\  {\bf 105}, 1413 (1957).


\bibitem{Goldhaber:1958nb} 
  M.~Goldhaber, L.~Grodzins and A.~W.~Sunyar,
    Phys.\ Rev.\  {\bf 109}, 1015 (1958).
    

\bibitem{Frauenfelder:1957na} 
  H.~Frauenfelder et al.,
    Phys.\ Rev.\  {\bf 106}, 386 (1957).
  
  

\bibitem{Page:1957zza} 
L.~A.~Page and M.~Heinberg, Phys.\ Rev.\  {\bf 106}, 1220 (1957).
    
    

\bibitem{Sudarshan:1958vf} 
  E.~C.~G.~Sudarshan and R.~E.~Marshak, Phys.\ Rev.\  {\bf 109}, 1860 (1958).

\bibitem{Feynman:1958ty} 
  R.~P.~Feynman and M.~Gell-Mann,
    Phys.\ Rev.\  {\bf 109}, 193 (1958).
  

\bibitem{Sakurai:1958zz} 
  J.~J.~Sakurai,
    Nuovo Cim.\  {\bf 7}, 649 (1958).
  

\bibitem{Salam:1957st}
A.~Salam,
Nuovo Cim. \textbf{5}, 299-301 (1957).


\bibitem{Landau:1957tp}
L.~D.~Landau,
Nucl. Phys. \textbf{3}, 127-131 (1957).

\bibitem{Lee:1957qr}
T.~D.~Lee and C.~N.~Yang,
Phys. Rev. \textbf{105}, 1671-1675 (1957).

\bibitem{Cabibbo:1963yz} 
  N.~Cabibbo, Phys.\ Rev.\ Lett.\  {\bf 10}, 531 (1963).
  
  

\bibitem{Weinberg:1967tq} 
  S.~Weinberg,
  Phys.\ Rev.\ Lett.\  {\bf 19}, 1264 (1967).

\bibitem{Salam:1968rm} 
  A.~Salam, Conf.\ Proc.\ C {\bf 680519}, 367 (1968).

\bibitem{Fatima:2018tzs}
A.~Fatima, M.~Sajjad Athar and S.~K.~Singh,
Phys. Rev. D \textbf{98}, 033005 (2018).

\bibitem{ParticleDataGroup:2020ssz}
P.~A.~Zyla \textit{et al.} [Particle Data Group],
PTEP \textbf{2020}, no.8, 083C01 (2020).

\bibitem{Ansari:2020xne}
V.~Ansari, M.~Sajjad Athar, H.~Haider, S.~K.~Singh and F.~Zaidi,
Phys. Rev. D \textbf{102}, no.11, 113007 (2020).
 

\bibitem{Callan:1969uq}
C.~G.~Callan, Jr. and D.~J.~Gross,
Phys. Rev. Lett. \textbf{22}, 156-159 (1969).
    
    

\bibitem{Albright:1974ts}
C.~H.~Albright and C.~Jarlskog,
Nucl. Phys. B \textbf{84}, 467-492 (1975).

\bibitem{Harland-Lang:2014zoa} 
  L.~A.~Harland-Lang, A.~D.~Martin, P. Motylinski and R.~S.~Thorne,
    Eur.\ Phys.\ J.\ C {\bf 75}, no. 5, 204 (2015).  

\bibitem{Eichten:1972bb}
T.~Eichten, H.~Faissner, S.~Kabe, W.~Krenz, J.~Von Krogh, J.~Morfin, K.~Schultze, J.~Lemonne, J.~Sacton and W.~Van Doninck, \textit{et al.}
Phys. Lett. B \textbf{40}, 593 (1972).

\bibitem{Erriquez:1977tr}
O.~Erriquez, M.~T.~Fogli Muciaccia, S.~Natali, S.~Nuzzo, A.~Halsteinslid, K.~Myklebost, A.~Rognebakke, O.~Skjeggestad, S.~Bonetti and D.~Cavalli, \textit{et al.}
Phys. Lett. B \textbf{70}, 383 (1977).

\bibitem{Erriquez:1978pg}
O.~Erriquez, M.~T.~Fogli-Muciaccia, S.~Natali, S.~Nuzzo, A.~Halsteinslid, C.~Jarlskog, K.~Myklebost, A.~Rognebakke, O.~Skjeggestad and B.~Tvedt, \textit{et al.}
Nucl. Phys. B \textbf{140}, 123 (1978).
 
 

\bibitem{Fanourakis:1980si}
G.~Fanourakis, L.~K.~Resvanis, G.~Grammatikakis, P.~Tsilimigras, A.~Vayaki, U.~Camerini, W.~F.~Fry, R.~J.~Loveless, J.~H.~Mapp and D.~D.~Reeder,
Phys. Rev. D \textbf{21}, 562 (1980).
 

\bibitem{SKAT:1989nel}
J.~Brunner \textit{et al.} [SKAT],
Z. Phys. C \textbf{45}, 551 (1990).
 

\bibitem{Kuzmin:2008zz}
K.~S.~Kuzmin and V.~A.~Naumov,
Phys. Atom. Nucl. \textbf{72}, 1501 (2009).
 

\bibitem{Wu:2013kla}
J.~J.~Wu and B.~S.~Zou,
Few Body Syst. \textbf{56}, 165 (2015).

\bibitem{Finjord:1975zy}
J.~Finjord and F.~Ravndal,
Nucl. Phys. B \textbf{106}, 228 (1976).
 

\bibitem{Bradford:2006yz} 
  R.~Bradford et al.,
    Nucl.\ Phys.\ Proc.\ Suppl.\  {\bf 159}, 127 (2006). 
    
    

\bibitem{Ammosov:1986jn}
V.~V.~Ammosov, V.~A.~Gapienko, G.~S.~Gapienko, A.~G.~Denisov, V.~I.~Klyukhin, V.~I.~Koreshev, P.~V.~Pitukhin, V.~I.~Sirotenko, Z.~U.~Usubov and V.~G.~Zaets, \textit{et al.}
Z. Phys. C \textbf{36}, 377 (1987).

\bibitem{Ammosov:1986xv}
V.~V.~Ammosov, A.~E.~Asratian, V.~A.~Gapienko, G.~S.~Gapienko, P.~A.~Gorichev, A.~G.~Denisov, V.~G.~Zaets, V.~I.~Klyukhin, V.~I.~Koreshev and S.~P.~Kruchinin, \textit{et al.}
JETP Lett. \textbf{43}, 716 (1986).


\bibitem{Nakamura:2015rta}
S.~X.~Nakamura, H.~Kamano and T.~Sato,
Phys. Rev. D \textbf{92}, 074024 (2015).
 

\bibitem{Hernandez:2007qq}
E.~Hernandez, J.~Nieves and M.~Valverde,
Phys. Rev. D \textbf{76}, 033005 (2007).

\bibitem{Gonzalez-Jimenez:2016qqq}
R.~Gonz\'alez-Jim\'enez, N.~Jachowicz, K.~Niewczas, J.~Nys, V.~Pandey, T.~Van Cuyck and N.~Van Dessel,
Phys. Rev. D \textbf{95},  113007 (2017).

\bibitem{Lalakulich:2010ss}
O.~Lalakulich, T.~Leitner, O.~Buss and U.~Mosel,
Phys. Rev. D \textbf{82}, 093001 (2010).

\bibitem{Wilkinson:2014yfa}
C.~Wilkinson, P.~Rodrigues, S.~Cartwright, L.~Thompson and K.~McFarland,
Phys. Rev. D \textbf{90}, 112017 (2014).

\bibitem{Rodrigues:2016xjj}
P.~Rodrigues, C.~Wilkinson and K.~McFarland,
Eur. Phys. J. C \textbf{76}, 474 (2016).


\bibitem{Tiator:2011pw}
L.~Tiator, D.~Drechsel, S.~S.~Kamalov and M.~Vanderhaeghen,
Eur. Phys. J. ST \textbf{198}, 141-170 (2011).

\bibitem{Radecky:1981fn}
G.~M.~Radecky, V.~E.~Barnes, D.~D.~Carmony, A.~F.~Garfinkel, M.~Derrick, E.~Fernandez, L.~Hyman, G.~Levman, D.~Koetke and B.~Musgrave, \textit{et al.}
Phys. Rev. D \textbf{25}, 1161 (1982)
[erratum: Phys. Rev. D \textbf{26}, 3297 (1982)].

\bibitem{Kitagaki:1986ct}
T.~Kitagaki, H.~Yuta, S.~Tanaka, A.~Yamaguchi, K.~Abe, K.~Hasegawa, K.~Tamai, S.~Kunori, Y.~Otani and H.~Hayano, \textit{et al.}
Phys. Rev. D \textbf{34}, 2554 (1986).


\bibitem{Gluck:2007ck}
M.~Gluck, P.~Jimenez-Delgado and E.~Reya,
Eur. Phys. J. C \textbf{53}, 355-366 (2008).

\bibitem{Nadolsky:2008zw}
P.~M.~Nadolsky, H.~L.~Lai, Q.~H.~Cao, J.~Huston, J.~Pumplin, D.~Stump, W.~K.~Tung and C.~P.~Yuan,
Phys. Rev. D \textbf{78}, 013004 (2008).

\bibitem{Martin:2009iq}
A.~D.~Martin, W.~J.~Stirling, R.~S.~Thorne and G.~Watt,
Eur. Phys. J. C \textbf{63}, 189-285 (2009).
 

\bibitem{Dulat:2015mca}
S.~Dulat, T.~J.~Hou, J.~Gao, M.~Guzzi, J.~Huston, P.~Nadolsky, J.~Pumplin, C.~Schmidt, D.~Stump and C.~P.~Yuan,
Phys. Rev. D \textbf{93}, no.3, 033006 (2016).
 

\bibitem{MicroBooNE:2023ubu}
P.~Abratenko \textit{et al.} [MicroBooNE],
Phys. Rev. Lett. \textbf{132}, no.15, 151801 (2024)
doi:10.1103/PhysRevLett.132.151801
[arXiv:2305.16249 [hep-ex]].

\bibitem{CHARM:1981zpm}
M.~Jonker \textit{et al.} [CHARM],
Phys. Lett. B \textbf{102}, 67 (1981).
 

\bibitem{Vermaseren:2005qc}
J.~A.~M.~Vermaseren, A.~Vogt and S.~Moch,
Nucl. Phys. B \textbf{724}, 3-182 (2005).

\bibitem{Moch:2008fj}
S.~Moch, J.~A.~M.~Vermaseren and A.~Vogt,
Nucl. Phys. B \textbf{813}, 220-258 (2009).


\bibitem{Zaidi:2019asc}
F.~Zaidi, H.~Haider, M.~Sajjad Athar, S.~K.~Singh and I.~Ruiz Simo,
Phys. Rev. D \textbf{101}, no.3, 033001 (2020).

\bibitem{CMS:2018wjk}
A.~M.~Sirunyan \textit{et al.} [CMS],
Phys. Rev. D \textbf{97}, 072010 (2018)
doi:10.1103/PhysRevD.97.072010
[arXiv:1802.04867 [hep-ex]].


\bibitem{Belle:2018ttu}
Y.~Guan \textit{et al.} [Belle],
Phys. Rev. Lett. \textbf{122},  042001 (2019)
doi:10.1103/PhysRevLett.122.042001
[arXiv:1808.05000 [hep-ex]].

\bibitem{LHCb:2013hzx}
R.~Aaij \textit{et al.} [LHCb],
Phys. Lett. B \textbf{724}, 27-35 (2013).




\end{thebibliography}
\end{document}